\documentclass[aps,prd,nofootinbib,aps,tightenlines,preprintnumbers,notitlepage,longbibliography,superscriptaddress]{revtex4-2}
\usepackage{graphicx}
\usepackage{amsmath}
\usepackage{amsfonts}
\usepackage[colorlinks=true,citecolor=blue,urlcolor=cyan,linkcolor=black]{hyperref}
\usepackage{appendix}
\usepackage{setspace}

\usepackage{bm}
\usepackage{amsmath}
\usepackage{amsfonts}

\usepackage{tabularx}
\newcolumntype{C}{>{\centering\arraybackslash}X}
\newcolumntype{R}{>{\raggedleft\arraybackslash}X}

\def\x{{\bf x}}
\def\y{{\bf y}}
\def\k{{\bf k}}
\def\K{{\bf K}}
\def\r{{\bf r}}
\def\l{{\bf l}}
\def\L{{\bf L}}
\def\hk{{\bf \hat k}}
\def\hx{{\bf \hat x}}

\def\hv{{\hat v}}
\def\th{{\bm\theta}}
\def\Th{{\bm\Theta}}
\def\hP{{\widehat P}}

\def\tB{{\widetilde B}}
\def\tS{{\widetilde S}}
\def\tT{{\widetilde T}}
\def\trho{{\widetilde\rho}}
\def\tphi{{\widetilde\phi}}
\def\fnl{f_{\rm NL}}
\def\bn{\bar n}
\def\bN{{\bar N}}
\def\nn{\nonumber}
\def\N{{\mathcal N}}

\usepackage[usenames, dvipsnames]{color}
\usepackage[normalem]{ulem}
\usepackage{xcolor}

\def\th{{\bm\theta}}

\newcommand{\be}{\begin{eqnarray}}

\newcommand{\ee}{\end{eqnarray}}

\definecolor{colorA}{HTML}{1E90FF}
\definecolor{colorB}{HTML}{228B22}
\definecolor{colorC}{HTML}{FF7F00}
\definecolor{colorD}{HTML}{4B0082}
\definecolor{colorE}{HTML}{B22222}

\definecolor{lgreen}{HTML}{32CD32}
\definecolor{lgray}{HTML}{D3D3D3}
\definecolor{dblue}{HTML}{1E90FF}
\definecolor{dblue}{HTML}{1E90FF}
\definecolor{orange}{HTML}{FF4500}
\definecolor{indigo}{HTML}{4B0082}

\definecolor{teal}{HTML}{008080}
\definecolor{firebrick}{HTML}{B22222}
\definecolor{salmon}{HTML}{FA8072}
\definecolor{darkgreen}{HTML}{006400}

\usepackage{multirow}
\usepackage{todonotes}

\newcommand{\aj}{Astron. J.}

\newcommand{\mnras}{Mon. Not. R. Astron. Soc.}
\newcommand{\jcap}{J. Cosm. Astropart. Phys.}
\newcommand{\pasp}{Publ. Astron. Soc. Pac.}

\newcommand{\araa}{Ann. Rew. of Astron. and Astrophys.}

\newcommand{\perimeter}{Perimeter Institute for Theoretical Physics, 31 Caroline St N, Waterloo, ON N2L 2Y5, Canada}

\newcommand{\berkeleya}{Lawrence Berkeley National Laboratory, One Cyclotron Road, Berkeley, CA 94720, USA}
\newcommand{\berkeleyb}{Berkeley Center for Cosmological Physics, Department of Physics, University of California, Berkeley, CA 94720, USA}

\graphicspath{ {plots/} }

\begin{document}

\title{Velocity Reconstruction from KSZ: Measuring $f_{NL}$ with ACT and DESILS}

\author{Selim~C.~Hotinli}
\affiliation{\perimeter}

\author{Kendrick~M.~Smith}
\affiliation{\perimeter}

\author{Simone~Ferraro}
\affiliation{\berkeleya}
\affiliation{\berkeleyb}

\begin{abstract}

The kinetic Sunyaev-Zel'dovich (kSZ) effect offers an indirect way to reconstruct large-scale cosmic velocities, by correlating high-resolution CMB temperature maps with galaxy surveys. 
In this work, we present the first three-dimensional reconstruction of the large-scale velocity field using a photometric galaxy survey, using data from the DESI Legacy Imaging Surveys (DESILS) and the Atacama Cosmology Telescope (ACT) DR5. 
We detect an $11.7\sigma$ correlation between our velocity reconstruction and the galaxy field, using only DESILS LRGs in the northern Galactic hemisphere.
We find that the overall amplitude of the kSZ-induced correlation is low relative to a halo model prediction ($b_v = 0.45^{+0.06}_{-0.05}$), in agreement with previous results which find high feedback and smoothed gas profiles near massive galaxies.
We use this measurement to place new constraints on local-type primordial non-Gaussianity (PNG), obtaining $f_{\rm NL}\!=\!-39^{+40}_{-33}$. This represents the most stringent $f_{\rm NL}$ constraint from kSZ velocity-based analyses to date. 
We validate our findings through extensive null tests, including tests for CMB foregrounds based on comparing 90 and 150 GHz CMB data.

\end{abstract} 

\maketitle

\section{Introduction}

Understanding the large-scale velocity field of the Universe is crucial for testing cosmological models and probing the matter distribution~\citep[see~e.g.][for reviews]{Hahn:2014lca,Koda:2013eya,Peacock:1996wd,Strauss:1995fz,Bertschinger:1993xt}. The velocity field is directly linked to structure growth, but direct measurements are notoriously difficult~\citep[e.g.][]{Courtois:2022mxo, Jasche:2018oym, Said:2020epb, Boruah:2019icj,Stiskalek:2025xaw} due to the lack of a universal tracer for cosmic motions. 

A promising alternative is the kinetic Sunyaev-Zel'dovich (kSZ) effect, a secondary anisotropy in the cosmic microwave background (CMB) caused by the Doppler shifting of photons scattering off moving free electrons~\citep{1980ARA&A..18..537S, 1980MNRAS.190..413S, 1970CoASP...2...66S}. 
Recently it has been shown that the kSZ effect can be used to reconstruct cosmic velocities, by cross-correlating CMB temperature maps with large-scale structure tracers, such as galaxies~\citetext{see~Refs.~\citealp{Deutsch:2017cja, Smith:2018bpn, Giri:2020pkk, Cayuso:2021ljq} for major developments in methods; and e.g.~Refs.~\citealp{Munchmeyer:2018eey,Cayuso:2019hen,Hotinli2019,Hotinli:2018yyc,Hotinli:2020csk,Hotinli:2020ntd,Hotinli:2021hih,AnilKumar:2022flx,Kumar:2022bly} for forecasts on a range of cosmological parameters}.  

One exciting near-term application of kSZ velocity reconstruction will be constraining primordial non-Gaussianity (PNG) of the local type~\cite{Munchmeyer:2018eey}, defined by a non-Gaussian primordial potential $\Phi$ of the form~\citep{Komatsu:2001rj} 
\begin{equation} 
\Phi(\x) = \phi(\x) + f_{\rm NL} \big( \phi(\x)^2 - \langle \phi^2 \rangle \big) \,,
\end{equation} 
where $\phi$ is a Gaussian field and $ f_{\rm NL} $ quantifies deviations from Gaussianity.
PNG, characterized by the $f_{\rm NL}$ parameter, quantifies deviations from Gaussian initial conditions and can reveal signatures of early-universe physics~\citep[e.g.][]{Verde:1999ij,Maldacena:2002vr,Bartolo:2004if}. 

The PNG imprints distinctive signatures on large-scale structure, including a scale-dependent bias in the clustering of tracers such as galaxies~\citep{Dalal:2007cu, Slosar:2008hx}. 
This effect is strongest for massive halos, grows at lower redshifts, and provides a powerful probe of non-Gaussian initial conditions using galaxy surveys like DESI~\citep{DESI:2016fyo,DESI:2022xcl, Chaussidon:2024qni}, Euclid~\citep{Amendola:2016saw}, LSST~\citep{LSSTScience:2009jmu} and SPHEREx~\citep{SPHEREx:2014bgr,SPHEREx:2016vbo,SPHEREx:2018xfm,2020SPIE11443E..0IC}.

While the most stringent constraints on \( f_{\rm NL} \) have traditionally come from CMB bispectrum measurements (e.g., \textit{Planck}’s \( f_{\rm NL} = 0.8 \pm 5.0 \)~\citep{Planck:2019kim}), large-scale structure probes offer a complementary approach by exploiting scale-dependent clustering signatures~\citep[e.g.][]{Andrews:2022nvv,DAmico:2022osl,DAmico:2022gki,Cabass:2022ymb,eBOSS:2021owp,Kurita:2023qku, Chaussidon:2024qni}. When combined with galaxy survey data, reconstructed velocities can help constrain the scale-dependent effects of local PNG on galaxy fluctuations through the galaxy-velocity correlation~\citep[e.g.][]{Munchmeyer:2018eey, AnilKumar:2022flx}. 
This is because PNG introduces the largest imprints on very large scales, where velocities are more sensitive than galaxy clustering alone. 

Previously, Ref.~\citep{Lague:2024czc} used the twelfth data release of the Sloan Digital Sky Survey
Baryon Oscillation Spectroscopic Survey (SDSS BOSS DR12~\citep{BOSS:2016wmc, 2013AJ....145...10D}), together with CMB temperature maps from ACT-DR5+\textit{Planck}~\citep{Naess:2020wgi,Planck:2018nkj} to perform a kSZ-based three-dimensional velocity reconstruction and measure $f_{\rm NL}\!=\!-90^{+210}_{-350}$ and detect velocity-galaxy cross-correlation at $7.2\sigma$. 

Concurrently, two complementary `light-cone' (two- or two-plus-one-dimensional) reconstructions have been performed on the data~\citep{McCarthy:2024nik,Krywonos:2024mpb,Bloch:2024kpn}. 
Refs.~\citep{Krywonos:2024mpb,Bloch:2024kpn} used \textit{Planck} PR3~\cite{PlanckCollaboration2020} \texttt{SMICA} and \texttt{Commander} component-separated temperature maps~\cite{Ade2014} and the unWISE~\cite{2010AJ....140.1868W, Mainzer2014,Schlafly_2019} `blue sample'~\cite{Krolewski2020} to constrain PNG within $-220<f_{\rm NL}<136$ at 68-percent confidence. 
Ref.~\citep{McCarthy:2024nik} combined the large-scale {\em spectroscopic} galaxy field from SDSS, the small-scale photometric galaxy field from the DESI Legacy Imaging Surveys (DESI-LS)~\citep{2019AJ....157..168D}, and high-resolution component separated ACT maps~\citep{ACT:2023wcq}, and detected the velocity field at $3.8\sigma$.

So far, no three-dimensional kSZ velocity reconstruction has used a galaxy sample with purely photometric redshifts.
Photometric surveys offer access to significantly larger cosmic volumes {and lower shot noise} than spectroscopic counterparts, but at the cost of increased redshift uncertainties, which can reduce sensitivity to radial velocity modes. 

In this work, we present the first three-dimensional kSZ velocity reconstruction using only a photometric galaxy survey. 
We use the galaxy data from DESI-LS~\citep{2019AJ....157..168D} and CMB temperature maps from ACT-DR5+\textit{Planck}~\citep{Naess:2020wgi,Planck:2018nkj}. We apply a quadratic estimator approach as introduced in Ref.~\citep{Smith:2018bpn} to extract the large-scale velocity field and measure the galaxy-velocity cross-power spectrum $P_{gv}(k)$ at $11.7\sigma$ significance. 

This paper is similar in scope to \cite{LKM}, which also does kSZ velocity reconstruction with ACT $\times$ (DESILS LRGs).
One methodological difference is that \cite{LKM} represents $\hv_r(\x)$ in spherical polar coordinates (Healpix $N_{\rm side}=32$) $\times$ (20 redshift bins), whereas we use 3-d Cartesian coordinates.
The cross correlation $\langle \delta_g \hat v_r \rangle$ is represented differently, as a 20-by-20 matrix of angular power spectra $C_l^{zz'}$ in \cite{LKM}, and as a 3-d power spectrum $P_{gv}(k)$ here.
The two papers use different analysis choices and data selection, e.g.\ we use less ACT data (DR5 vs DR6) and less DESILS data (Northern Galactic hemisphere only, in order to minimize imaging systematics).
Nevertheless, the two papers obtain very similar results, including the main result that the small-scale galaxy electron power spectrum $P_{ge}(k)$ is significantly smaller than the halo model prediction ($b_v \sim 0.4$).
Our paper was submitted to the arXiv simultaneously with \cite{LKM}, and the two papers are highly complementary.

This detection provides direct evidence that kSZ-based velocity reconstructions are feasible even with photometric redshift uncertainties. We use our $P_{gv}(k)$ measurement to place new constraints on local-type PNG, finding $f_{\rm NL}\!=\!-39^{+40}_{-33}$ using reconstructed 3-dimensional velocities from the DESI-LS and ACT-DR5 data. This is the most stringent $f_{\rm NL}$ constraint to date from kSZ velocity reconstruction, complementing and extending constraints from galaxy clustering and the CMB bispectrum.
The main results of this paper were produced using the {\tt kszx} software package, publicly available at \url{https://github.com/kmsmith137/kszx}.

\section{Preliminaries}

We denote 3-d comoving locations by $\x$ (three-vector), and 2-d sky locations by $\th$ (unit three-vector).
The line-of-sight direction (a unit three-vector oriented away from the observer) is denoted $\hx$.

We denote the comoving distance to redshift $z$ by $\chi(z)$, and will sometimes implicitly change variable $\chi \leftrightarrow z$.
Our Fourier conventions are:
\begin{equation}
f(\x) = \int \frac{d^3\k}{(2\pi)^3} \, \tilde f(\k) e^{i\k\cdot\x}
 \hspace{1.5cm}
\tilde f(\k) = \int d^3\x \, f(\x) e^{-i\k\cdot\x}\,.
\end{equation}
The kSZ contribution to the CMB can be written as a line-of-sight integral \cite{Sunyaev1980}:
\begin{equation}
    T_{\rm kSZ}(\th) = \int d\chi \, K(\chi) \, 
    v_r(\chi\th) \, \delta_e(\chi\th)\,,
    \label{eq:tksz_line_of_sight}
\end{equation}
where $\chi$ is comoving distance, $\th$ is the line-of-sight direction (a unit three-vector), $\delta_e(\chi\th)$ is the electron overdensity and $v_r(\chi\th)$ is the radial velocity. 
The quantity $K(z)$ appearing in Eq.\ (\ref{eq:tksz_line_of_sight}) is defined by
\begin{equation}
K(z) \equiv -T_{\rm CMB} \, \sigma_T \, n_{e0} \, x_e(z) \, e^{-\tau(z)} \, (1+z)^2\,,
\label{eq:K_def}
\end{equation}
where $T_{\rm CMB}$ is the mean CMB temperature, $\sigma_T$ is the Thomson cross-section, $n_{e0}$ is the mean electron density today, $x_e(z)$ is the ionization fraction at redshift $z$ and $\tau(z)$ is the optical depth to redshift $z$.

In an $f_{\rm NL} \ne 0$ cosmology, galaxy bias acquires an extra term on large scales \cite{Dalal:2007cu, Slosar:2008hx}:
\begin{equation}
b(k) = b_g + 2 \delta_c (b_g-1) \frac{f_{\rm NL}}{\alpha(k,z)}
  \hspace{1.5cm} \mbox{where }
\alpha(k,z) = \frac{\delta_m(\k,z)}{\Phi(\k,z)} = \frac{2 k^2 T(k) D(z)}{3 \Omega_{m0} H_0^2}\,.
\label{eq:bg_fnl}
\end{equation}
Here, \( T(k) \) is the matter transfer function, \( D(z) \) the linear growth function, \( \Omega_{m0} \) is the present-day matter density, \( H_0 \) is the Hubble constant, and $\delta_c=1.68$ is the collapse threshold. 
This leads to an enhancement of clustering on large scales proportional to $(f_{NL}/k^2)$, since $\alpha(k,z) \propto k^2$ as $k \rightarrow 0$.

\section{High-level pipeline description}
\label{sec:high_level_pipeline}

In this section, we describe features of our kSZ velocity reconstruction pipeline which are not specific to the datasets under consideration (ACT and DESI-LS).

\subsection{The KSZ velocity reconstruction estimator $\hat v_r$}
\label{ssec:hvr}

The first step in our pipeline is the kSZ quadratic estimator $\hat v_r$, which reconstructs the large-scale velocity field (up to a multiplicative bias $b_v$). The inputs are the 3-d galaxy field $\delta_g(\x)$ and the 2-d CMB $T(\th)$, and  the output is a 3-d field $\hat v_r(\x)$ which reconstructs the velocity field on large scales.

To motivate the definition of $\hat v_r$ that we use in our pipeline, we start from the expression for $\hat v_r$ in \cite{Smith:2018bpn}, which applies to a simplified flat-sky ``snapshot'' geometry with no sky cut:
\begin{equation}
\hv_r(\k_L) \propto \int_{\k_S + (\l/\chi_*) = \k_L}
  \frac{P_{ge}(k_S)}{P_{gg}^{\rm tot}(k_S)} \frac{1}{C_l^{\rm tot}}
  \Big( \delta_g(\k_S) T_{\rm CMB}(\l) \Big)\,. \label{eq:vsch1}
\end{equation}
Here, $\chi_*$ is comoving distance to the galaxies (assumed constant in this simplified snapshot geometry), $C_\ell^{\rm tot}$ is the total CMB power spectrum (including foregrounds and noise), $P_{ge}(k)$ is the electron-galaxy correlation, and $P_{gg}^{\rm tot}(k)$ is the total galaxy power spectrum (including Poisson noise).
We have used the notation $\k_L, \k_S$ to emphasize that the LHS is usually evaluated at large scales $k_L \sim 10^{-2}$ Mpc$^{-1}$, and the integrand on the RHS peaks on small scales $k_S \sim 1$ Mpc$^{-1}$.

As written, Eq.\ (\ref{eq:vsch1}) involves the small-scale galaxy field $\delta_g(\k_S)$, which would be memory and compute intensive to store at high resolution ($k_S \sim 1$ Mpc$^{-1}$).
Following \cite{Lague:2024czc}, we will make the approximation $k_S \approx l/\chi_*$ inside the integral.
This is an excellent approximation in the squeezed limit $k_L \ll k_S$, and allows $\hat v_r$ to be written this way:
\begin{align}
\hv_r(\k_L) &\propto \int_{\k_S + (\l/\chi_*) = \k_L}
  \frac{P_{ge}(l/\chi_*)}{P_{gg}^{\rm tot}(l/\chi_*)} \frac{1}{C_l^{\rm tot}}
  \Big( \delta_g(\k_S) T_{\rm CMB}(\l) \Big)\,.
\end{align}
Or, switching from Fourier space to real space:
\begin{equation}
\hv_r(\x) = \sum_{i \in \rm gal} \tT(\th_i) \delta^3(\x-\x_i)\,, \label{eq:vsch2}
\end{equation}
where $\th_i$ is the angular location of the $i$-th galaxy, and the high-pass filtered CMB $\tT(\th)$ is defined by:
\begin{equation}
\tT(\th) = \int \frac{d^2\l}{(2\pi)^2} \, \tT(\l) e^{i\l\cdot\th}\,,
 \hspace{1.5cm}
\tT(\l) \equiv \frac{P_{ge}(l/\chi_*)}{P_{gg}(l/\chi_*)} \frac{1}{C_l^{\rm tot}} T_{\rm CMB}(\l)\,.
\label{eq:tT_snapshot}
\end{equation}
The point of writing $\hat v_r$ in the form (\ref{eq:vsch2}) is to reduce computational requirements, by avoiding the need to store high-resolution 3-d maps such as $\delta_g(\k_S)$.
In code, the field $\hv_r(\x)$ can be represented as either a weighted catalog (with per-galaxy weight $\smash{\tT(\theta_i)}$, or a 3-d field in a low-resolution pixelization (after ``gridding'' the weighted catalog).

So far, we have used a simplified snapshot geometry (following \cite{Smith:2018bpn}).
Next, we will generalize to a lightcone geometry with a sky cut.
We will also introduce two weight functions: a CMB pixel weight function $W_{\rm CMB}(\th)$, and a per-galaxy weight $W_i^v$.
In our ACT-DESILS pipeline, $W_{\rm CMB}(\th)$ will be used to downweight regions of high noise or large CMB foreground emission, and $W_i^v$ will be used to downweight galaxies with large photo-$z$ errors (see \S\ref{sec:data}).

First, we define the all-sky filtered CMB $\tT_{lm}$ by:
\begin{equation}
\tT(\th) = \sum_{lm} \tT_{lm} Y_{lm}(\th)\,,
 \hspace{1cm}
\tT_{lm} = F_l \, \int d^2\th \, W_{\rm CMB}(\th) \, T_{\rm CMB}(\th) \, Y_{lm}^*(\th)\,,
  \hspace{1cm}
F_l \equiv \frac{P_{ge}(l/\chi_*)}{P_{gg}(l/\chi_*)} \frac{b_l}{C_l^{\rm tot}} \,.
\label{eq:tT_allsky}
\end{equation}
This is just the all-sky version of Eq.\ (\ref{eq:tT_snapshot}), with the pixel weight function $W_{\rm CMB}(\th)$ included by hand.
Then, we define $\hat v_r$ by:
\begin{equation}
\hv_r(\x) = \sum_{i\in \rm gal} W_i^v \, \tT(\th_i) \delta^3(\x-\x_i)\,,
\label{eq:hvr_no_ms}
\end{equation}
following Eq.\ (\ref{eq:vsch2}), with the per-galaxy weighting $W_i^v$ included by hand.
In implementation, we only keep terms in the sum (\ref{eq:hvr_no_ms}) where $W_{\rm CMB}(\th) > 0$. 

Empirically, we find that in order to mitigate CMB foregrounds, the above construction must be modified, by subtracting the mean temperature $\tT(\theta_i)$ within redshift bins of size $\Delta z \sim0.1$.
More precisely, for each redshift bin $b$, let $\tT_b$ be the mean value of $\tT(\th_i)$ over all galaxies in the bin:
\begin{equation}
\tT_b \equiv \frac{\sum_{i\in b} W_i^v \tT(\th_i)}{\sum_{i\in b} W_i^v}\,.
\end{equation}
Then, we define a ``mean-subtracted'' velocity reconstruction $\hat v_r(\x)$ by:
\begin{equation}
\hv_r(\x) = \sum_b \sum_{i\in b} W_i^v \, \big( \tT(\th_i) - \tT_b \big) \delta^3(\x-\x_i)\,. \label{eq:hvr_ms}
\end{equation}
To motivate this mean-subtraction step intuitively, consider a CMB foreground which is correlated with the galaxy survey (such as tSZ, CIB, etc.)
On average, such a foreground will make an additive contribution to the value of $\tT(\th_i)$ at the locations of the galaxies, in a way which is slowly varying in redshift and independent of radial velocity $v_r$.
If $\hv_r$ is defined without mean subtraction (Eq.\ (\ref{eq:hvr_no_ms})) then such a foreground makes a systematic contribution to $\hv_r$ (more precisely, $\hv_r(\x)$ gets a term which is proportional to $\delta_g(\x)$).
On the other hand, if $\hv_r$ is defined with mean subtraction (Eq.\ (\ref{eq:hvr_ms})) then the CMB foreground cancels on average.

A subtlety: the mean subtraction step in Eq.\ (\ref{eq:hvr_ms}) can potentially bias power spectrum estimates, such as the cross correlation $P_{gv_r}(k)$ with a galaxy field.
At field level, mean subtraction is the same thing as ``projecting out modes'' from the 3-d field $\hat v_r(\x)$.
More precisely, any 3-d mode which is slowly varying in redshift, and constant in angular coordinates $\th$, is projected out.
This is very similar to the ``radial integral constraint'' for a galaxy field \cite{deMattia:2019vdg}, which removes modes and can bias the power spectrum.
In \S\ref{ssec:surr2} we will explain our method for removing this bias, and accounting for mode removal when assigning error bars.

The velocity reconstruction $\hat v_r(\x)$ is a biased reconstruction of the radial velocity field, but the overall normalization is nontrivial to derive.
In Appendix \ref{app:hvr_norm} we show that the normalization can be written (with some approximations that are explained in the appendix) as:
\begin{equation}
\big\langle \hat v_r(\x) \big\rangle = b_v \, \bar n_v(\x) \, B(\x) \, v_r^{\rm true}(\x)\,,
\label{eq:vrec_bias}
\end{equation}
where $\bn_v(x)$ is the 3-d galaxy number density including the per-object weighting $W_i^v$, defined formally by:
\begin{equation}
n_v(\x) \equiv \left\langle \sum_i W_i^v \delta^3(\x-\x_i) \right\rangle\,,
\label{eq:nv_def}
\end{equation}
and the function $B(\x)$ is defined by:
\begin{equation}
B(\x) \equiv  W_{\rm CMB}(\th) \, B_0(\chi)\,,
 \hspace{1.5cm}
B_0(\chi) \equiv \frac{K(\chi)}{\chi^2} \int \frac{d^2\L}{(2\pi)^2} \,
   b_L \, F_L \, P_{ge}^{\rm fid}(k,\chi)_{k=L/\chi}\,,
      \label{eq:B_def}
\end{equation}
where $b_L$ is the CMB beam, and the quantities $K(\chi)$, $F_L$ were defined in Eqs.\ (\ref{eq:K_def}), (\ref{eq:tT_allsky}).

Finally, the ``kSZ velocity bias'' $b_v$ in Eq.\ (\ref{eq:vrec_bias}) is an overall prefactor that is roughly the ratio between the true galaxy-electron power spectrum $P_{ge}^{\rm true}(k)$ and the fiducial galaxy-electron power spectrum $P_{ge}^{\rm fid}(k)$ used to define the estimator in Eq.\ (\ref{eq:tT_allsky}).
The precise definition is:
\begin{equation}
b_v(\chi) = \frac{
\int (d^2\L/(2\pi)^2)) \, b_L F_L P_{ge}^{\rm true}(k,\chi)_{k=L/\chi}}{\int (d^2\L/(2\pi)^2)) \, b_L F_L P_{ge}^{\rm fid}(k,\chi)_{k=L/\chi}}\,.
\label{eq:bv_def}
\end{equation}
In our pipeline, we will treat $b_v$ as a nuisance parameter and marginalize it (neglecting redshift dependence).
The presence of such a nuisance parameter is a general feature of kSZ velocity reconstruction \cite{Smith:2018bpn}.

Finally, we comment on the normalization of $\hv_r(\x)$.
Our definition of $\hv_r(\x)$ in Eqs.\ (\ref{eq:tT_allsky}), (\ref{eq:hvr_no_ms}) uses an arbitrary normalization -- e.g.\ $\hv_r(\x)$ has units Mpc$^{-3}$ $\mu$K$^{-1}$, whereas the physical radial velocity field $v_r(\x)$ is dimensionless.
The function $B(\x)$ defined in Eq.\ (\ref{eq:B_def}) keeps track of the normalization which relates $\hv_r(\x)$ and $v_r(\x)$.
In the next section, we'll define a power spectrum estimator $\hP_{gv}(k)$ which is physically normalized (via the quantity $\N_{gv}$, appearing in Eq.\ (\ref{eq:hpgv_def}) and defined in Appendix \ref{app:hvr_norm}, which contains one power of $B(\x)$.)

\subsection{The power spectrum estimator $P_{gv}$(k)}
\label{ssec:pgv}

We define the three-dimensional large-scale galaxy field $\rho_g(\x)$ through the `\textit{galaxies-minus-randoms}' prescription:
\begin{equation}
\rho_g(\x) \equiv\!\!
 \bigg( \sum_{i\in \rm gal}\!W_i^g \, \delta^3(\x-\x_i) \bigg)
  \!-\!\frac{N_g}{N_r} 
    \bigg( \sum_{j\in \rm rand}\!\!\!W_j^g \, \delta^3(\x-\x_j) \bigg)\,,
    \label{eq:rhog_def}
\end{equation}
where $W_i^g$ is a per-galaxy weight (e.g.\ FKP), and  $N_g$ and $N_r$ are the number of galaxies in the data catalog and randoms, respectively.
Note that we distinguish notationally between the weights $W_i^g$ and $W_i^v$ used in the galaxy field $\rho_g$ and velocity reconstruction $\hat v_r$ (in Eqs.\ (\ref{eq:hvr_ms}), (\ref{eq:rhog_def})).
However, in our DESILS pipeline, we choose the weights $W_i^g = W_i^v$ to be equal (Eq.\ (\ref{eq:wfkp_photoz})).

We define the power spectrum estimators $\hP_{gg}(k)$ and $\hP_{gv}(k)$ by:
\begin{equation}
\hP_{gg}(k) \equiv \frac{1}{\N_{gg}} \int \frac{d\Omega_k}{4\pi} \, 
 \rho_g^*(\k) \, \rho_g(\k)\,,
\hspace{1.5cm}
\hP_{gv}(k) \equiv \frac{3}{\N_{gv}} \int \frac{d\Omega_k}{4\pi} \, 
 \rho_g^*(\k) \, \hv_1(\k)\,,
 \label{eq:hpgv_def}
\end{equation}
where the field $\hv_1(\k)$ is defined in terms of the reconstructed radial velocity $\hv_r(\x)$ as
\begin{equation}
\hv_1(\k) \equiv \int_{\x} (i \hk_j \cdot\hx_j) \, \hv_r(\x) \, e^{-i\k\cdot\x}\,.
\end{equation}
The estimator $\hat P_{gv}(k)$ defined in Eq.\ (\ref{eq:hpgv_def}) is the standard ``dipolar'' power spectrum estimator, for cross-correlating a scalar field $\rho_g$ with an ``$\ell=1$'' field $\hat v_r$, in a wide-angle geometry \cite{Scoccimarro:2015bla,Hand:2017irw}.

The normalizing prefactors $\N_{gg}$, $\N_{gv}$ in Eq.\ (\ref{eq:hpgv_def}) are defined in Appendix \ref{app:power_spectrum_normalization}, and are intended to approximately correct for the survey geometry.
This is an approximation, since the true effect of the survey geometry is $k$-dependent (and also slightly non-diagonal in $k$: the estimator $\smash{\hP_{gv}(k)}$ receives contributions from $P_{gv}(k')$ at wavenumbers $k'\ne k$).
Therefore, our power spectrum estimator $\smash{\hP_{gv}(k)}$ is not a perfectly unbiased estimator of $P_{gv}(k)$.

This is an issue, since the main goal of this paper is to compare the values of $\hP_{gv}(k)$ to a two-parameter model for $P_{gv}(k)$:
\begin{equation}
P_{gv}(k) = b_v \left( b_g + f_{\rm NL} \frac{2\delta_c (b_g-1)}{\alpha(k,z)} \right) \frac{faH}{k} P_{\rm lin}(k)\,.
\end{equation}
If $\hP_{gv}(k)$ is not strictly unbiased, then comparing to a model $P_{gv}(k)$ is not straightforward.

One approach would be to modify the definition of the estimator $\hP_{gv}(k)$ to make it strictly unbiased, by computing and deconvolving the effect of the survey geometry.
In future work, we plan to pursue this approach, which is appealing but nontrivial (e.g.\ in \cite{Lague:2024czc} it is argued that existing approximations break down in the lowest $k$-bin of SDSS, and improved approximations are needed).

In this paper, we will use a different approach: rather than deconvolving the estimator $\hP_{gv}(k)$, we will convolve the model $P_{gv}(k)$ with the survey geometry.
That is, we leave the estimator $\smash{\hP_{gv}(k)}$ in the simple form defined by Eq.\ (\ref{eq:hpgv_def}), but whenever we compare the estimator to a model, we compute the estimator response $\smash{\langle \hat P_{gv}(k) \rangle}$ to the model $P_{gv}(k)$, in a way which precisely accounts for the survey geometry.
We'll do this using the machinery of ``surrogate fields'', which we describe next.

\subsection{Surrogate fields, part 1: galaxy field}
\label{ssec:surr1}

In this section and the next, we introduce {\em surrogate fields}, which will play a central role in our pipeline.
A ``surrogate field'' is a simplified random field that preserves the exact field-level covariance $\langle \delta(\x) \delta(\x') \rangle$ of a more complicated random field (including survey geometry, light-cone evolution, and photometric redshift uncertainties).
This is in contrast to a ``mock'', which is a realistic simulation that captures higher-order statistics (beyond the two-point function).
We will argue that for purposes of this paper, where analysis is limited to large scales ($k \lesssim 0.02$ Mpc$^{-1}$), it suffices to use simplified surrogate fields in place of mocks.

It is easiest to describe surrogate fields by example.
Consider the galaxy field $\rho_g(\x)$, which can be modeled on large scales as:
\begin{equation}
\rho_g(\k,z) = \bar n_g(z) \big[ 1 + \delta_G(\k,z) \big]
+ (\mbox{Poisson noise})\,,
\label{eq:deltag_model1}
\end{equation}
where $\delta_G$ is the Gaussian field:
\begin{equation}
\delta_G(\k,z) \equiv
\left( b_g(z) + f_{\rm NL} \frac{2\delta_c (b_g(z)-1)}{\alpha(k,z)} \right)
\delta_{\rm lin}(\k,z)\,.
\label{eq:deltag_model2}
\end{equation}
In this context, a ``mock'' is a catalog of objects such that the corresponding sum of delta functions $\sum_i W_i^g \delta^3(\x-\x_i)$ is described on large scales by the model in Eq.\ (\ref{eq:deltag_model1}).
A ``surrogate field'' can be any random field $S_g(\x)$ (not necessarily a sum of delta functions) whose field-level covariance $\langle S_g(\x) S_g(\x') \rangle$ matches the covariance $\langle \rho_g(\x) \rho_g(\x') \rangle$.
Note that $\rho_g(\x)$ is defined with the survey mask applied, so the surrogate field must fully match the survey geometry, in order to match the field-level covariance.

We construct such a surrogate field $S_g(\x)$ as follows.
Our goal is to write down the simplest possible random field that has the same covariance as $\rho_g(\x)$, assuming the clustering model (\ref{eq:deltag_model1}).
Since this model consists of two terms (a signal term and a Poisson noise term), our surrogate field will consist of two terms:
\begin{equation}
S_g(\x) = S_g^{\rm sig}(\x) + S_g^{\rm noise}(\x) \,.
  \label{eq:Sg_def}
\end{equation}
We define the signal term $S_g^{\rm sig}(\x)$ as follows.
In each Monte Carlo iteration, we randomly simulate a {\em Gaussian} field $\delta_G(\x)$ directly from its definition in Eq.\ (\ref{eq:deltag_model2}).
Then, we define $S_g^{\rm sig}(\x)$ by:
\begin{equation}
S_g^{\rm sig}(\x) \equiv \frac{N_g}{N_r} \sum_{j\in \rm rand} W_j^g \, 
 \delta_G(\x_j) \, \delta^3(\x-\x_j)\,.
 \label{eq:Sg_sig_def}
\end{equation}
Note that we sum over randoms, not galaxies!
The purpose of the random catalog is to capture the survey geometry and redshift dependence.

Similarly, we define the noise term $S_g^{\rm noise}(\x)$ by adding Gaussian white noise, whose variance is chosen to give the correct (Poisson) power spectrum $P_{gg}(k)$ at high $k$:
\begin{equation}
S_g^{\rm noise}(\x) = \frac{N_g}{N_r} \sum_{j\in \rm rand} W_j^g \,  \eta_j \, \delta^3(\x-\x_j)\,,
 \label{eq:Sg_noise_def}
\end{equation}
where $\eta_j$ is an independent (for each $j$) Gaussian random variable with variance:
\begin{equation}
\big\langle \eta_j^2 \big\rangle = 
 \frac{N_r}{N_g} - \big\langle \delta_G(z_j)^2 \big\rangle\,.
 \label{eq:sgg_eta}
\end{equation}
The surrogate field $S_g(\x)$ defined by Eqs. (\ref{eq:Sg_def})--(\ref{eq:sgg_eta}) has a covariance $\langle S_g(\x) S_g(\x') \rangle$ which agrees with the model in Eqs.\ (\ref{eq:deltag_model1}), (\ref{eq:deltag_model2}).
This intuitively plausible statement is proved rigorously in Appendix \ref{app:surrogates}.

Note that the surrogate field $S_g(\x)$ defined by Eqs. (\ref{eq:Sg_def})--(\ref{eq:sgg_eta}) does not really resemble a mock catalog.
The locations of the delta functions are unclustered (i.e.\ the sum in Eq.\ (\ref{eq:Sg_sig_def}) runs over randoms, not galaxies), and $S_g(\x)$ gets clustering power by ``painting'' the Gaussian field $\delta_G(\x)$ onto the randoms.
The surrogate field $S_g(\x)$ does not have a preferred sign (in contrast to a mock, which is a sum of delta functions).
Nevertheless, the surrogate field can be used in place of a mock, in situations where the covariance is important and higher-order correlation functions are not.
It may help to think of the surrogate field as a Monte Carlo representation of the field-level covariance, not a simulated dataset.

Why do we need surrogate fields?
In our pipeline, surrogate fields solve several problems at once:

\begin{itemize}

\item Computing estimator expectation values $\langle \hP_{gg}(k) \rangle$ or $\langle \hP_{gv}(k) \rangle$, given a model for the underlying random fields (as mentioned at the end of the previous section).
Since such expectation values only depend on the field-level covariance, they can be computed by taking an average over simulated surrogate fields (which have the correct covariance by construction).

\item
Estimating estimator covariance, for example $\mbox{Cov}(\hP_{gv}(k), \hP_{gv}(k'))$, by taking a Monte Carlo average over surrogates.
It is not obvious {\em a priori} that this is a good approximation, since the power spectrum covariance includes a ``disconnected'' contribution from the 4PCF, which may not be accurately captured by surrogate fields.
In this paper, we limit our analysis to large scales ($k \lesssim 0.02$ Mpc$^{-1}$), where large-scale structure fields are Gaussian, and the 4PCF is expected to be small.
In the context of kSZ velocity reconstruction, it has been shown in \cite{Giri:2020pkk} using full $N$-body simulations that $\hat v_r$ is a Gaussian field to a good approximation on such scales.

As a further check that surrogates accurately approximate power spectrum covariance on large scales, we compared $\smash{\mbox{Cov}(\hP_{gg}(k), \hP_{gg}(k'))}$ using surrogates and full mocks, in SDSS (not DESILS, since mocks are not available for DESILS).
We found that the surrogates accurately approximate both the diagonal and the off-diagonal covariance, for $k_{\rm max} = 0.02$ Mpc$^{-1}$.
The detailed comparison is given in Appendix \ref{sec:Appendix_SDSSvalidation}.

\item Parameter dependence of expectation values and covariance.
As a concrete example, for the galaxy field $\rho_g(\x)$ considered in this section, both the expectation value $\smash{\langle \hP_{gg}(k) \rangle}$ and the covariance $\smash{\mbox{Cov}(\hP_{gg}(k), \hP_{gg}(k'))}$ depend on the model parameter $\fnl$.
In principle, this $\fnl$ dependence can be determined by running surrogate simulations with different values of $\fnl$, where $S_g(\x)$ depends on $\fnl$ via the second term in Eq.\ (\ref{eq:deltag_model2}).

As an optimization, we note that $S_g(\x)$ depends linearly on $\fnl$:
\begin{equation}
S_g(\x) = S_g^{(0)}(\x) + \fnl \, S_g^{(1)}(\x)\,.
\end{equation}
In our pipeline, we generate the fields $S_g^{(0)}(\x)$, and $S_g^{(1)}(\x)$ separately in each Monte Carlo simulation, so that the full $\fnl$ dependence of $\smash{\langle \hP_{gg}(k) \rangle}$ and the covariance $\smash{\mbox{Cov}(\hP_{gg}(k), \hP_{gg}(k'))}$ can be computed, without running new simulations for each $\fnl$ value of interest.
A similar optimization is used in the next section, to compute the $b_v$ dependence of $\hat P_{gv}$ and its covariance.

\end{itemize}

\par\noindent
The primary advantage of surrogate fields lies in their ease of construction compared to realistic mocks. 
In particular, one challenge of working with DESILS is that the mocks are not publicly available, and making realistic DESILS mocks ``from scratch'' would be a daunting proposition.
On the other hand, surrogate simulations are fairly straightforward to implement, computationally inexpensive (a few CPU-minutes per simulation), and suffice for the large-scale analysis $(k \lesssim 0.02$ Mpc$^{-1}$) of this paper.

\subsection{Surrogate fields, part 2: KSZ velocity reconstruction}
\label{ssec:surr2}

In the previous section, we showed how to construct a surrogate field $S_g(\x)$ with the same field-level covariance as the galaxy density field $\rho_g(\x)$.
Analogously, in this section we will show how to construct a surrogate field $S_v(\x)$ with the same covariance as the kSZ velocity reconstruction $\hv_r(\x)$.

Our starting point is the following large-scale model for $\hv_r(\x)$, introduced previously in Eq.\ (\ref{eq:vrec_bias}):
\begin{equation}
\hv_r(\x) = b_v \, \bar n_v(\x) \, B(\x) \, v_r^{\rm true}(\x)
 + (\mbox{kSZ reconstruction noise})\,.
 \label{eq:hvr_model}
\end{equation}
Since this model consists of two terms (a signal term and a noise term), our surrogate field will consist of two terms:
\begin{equation}
S_v(\x) = S_v^{\rm sig}(\x) + S_v^{\rm noise}(\x) \,.
 \label{eq:Sv_sum}
\end{equation}
We define the signal term $S_v^{\rm sig}(\x)$ as follows.
In each Monte Carlo simulation, let $v_r^{\rm sim}(\x)$ be the (linear) radial velocity field.
Then we define $S_v^{\rm sig}(\x)$ by as a sum over randoms (not galaxies):
\begin{equation}
S_v^{\rm sig}(\x) \equiv
\frac{N_g}{N_r} \sum_{j\in\rm rand} 
  W_j^v B(\x_j) \, v_r^{\rm sim}(\x_j) \, \delta^3(\x-\x_j) \,,
  \label{eq:Sv_sig}
\end{equation}
where $B(\x)$ was defined in Eq.\ (\ref{eq:B_def}).

Next, we consider the noise term $S_v^{\rm noise}(\x)$.
By definition, kSZ reconstruction noise is the contribution to $\hat v_r$ that would arise if the CMB were statistically independent of the 3-d large-scale structure.
One approach to simulating reconstruction noise would be to simulate a random CMB realization (signal + noise) in each Monte Carlo realization.
This is nontrivial since it involves modeling details such as anisotropic noise and foreground correlations between 90 and 150 GHz.
We use a simpler approach: a bootstrap procedure based on the real ACT maps.
In each Monte Carlo realization, we choose a random size-$N_g$ subset $S$ of the random catalog $\{1,\cdots,N_r\}$.
Then we define a bootstrap estimate of the reconstruction noise by keeping only randoms in $S$:
\begin{equation}
S_v^{\rm noise}(\x) \equiv \sum_{j\in\rm rand} M_j \, W_j^v \, \tT(\th_j) \delta^3(\x-\x_j)
\hspace{1cm} \mbox{where }
M_j \equiv \begin{cases}
 1 & \mbox{ if $j \in S$} \\
 0 & \mbox{ if $j \not\in S$}
\end{cases}
\label{eq:Sv_noise}
\end{equation}
where $\tT(\th_j)$ is the high-pass-filtered ACT map (not a simulated ACT map), defined previously in Eq. (\ref{eq:tT_allsky}).

In Appendix \ref{app:hvr_norm}, we show that the surrogate field $S_v(\x)$ defined above correctly models the field-level covariance of $\hat v_r(\x)$.
In our pipeline, we will use surrogate fields to estimate $\smash{\langle \hP_{gv}(k) \rangle}$ and $\smash{\mbox{Cov}(\hP_{gv}(k), \hP_{gv}(k'))}$ by Monte Carlo, including dependence on the parameters $(\fnl, b_v)$.
We highlight the following aspects of our surrogate field construction:

\begin{itemize}

\item
The surrogate fields $S_g(\x)$ and $S_v(\x)$ will be correlated on large scales.
This is because the surrogate fields are constructed from Gaussian fields $\delta_G(\x)$ and $v_r^{\rm sim}(\x)$, which are highly correlated since they are generated from the same realization of the linear density field.
The large-scale correlation between $S_g(\x)$ and $S_v(\x)$ is important when using surrogate simulations to compute $\smash{\langle \hP_{gv}(k) \rangle}$, or sample variance contributions to $\smash{\mbox{Cov}(\hP_{gv}(k), \hP_{gv}(k'))}$.

\item 
We simulate surrogate fields $S_v(\x)$ for 90 and 150 GHz CMB maps.
The ``signal'' components $S_v^{\rm sig}(\x)$ are 100\% correlated, since the radial velocity $v_r^{\rm sim}(\x)$ is the same.
Less obviously, the ``noise'' components $S_v^{\rm sig}(\x)$ are also correlated, since the underlying CMB maps $\smash{\tT_{90}(\th})$ and $\smash{\tT_{150}(\th)}$ are correlated due to shared CMB and foregrounds.
The signal and noise correlations will be important later, when combining 90 and 150 GHz to get joint constraints.

\end{itemize}

One more detail: mode removal in surrogate fields.
Recall from \S\ref{ssec:hvr} that when we compute $\hv_r$ on data, we introduced a mean-subtraction step (Eq.\ (\ref{eq:hvr_no_ms})) in order to mitigate CMB foregrounds.
This effectively projects out modes in the 3-d field $\hv_r(\x)$.
To make our surrogate simulations consistent, we perform the same mean-subtraction step on the surrogate field $S_v(\x)$.
Then, when we use surrogate simulations to compute the mean $\smash{\hP_{gv}(k)}$ and assign error bars, the surrogate sims will account for biases due to mode removal in $\hv_r(\x)$.

Similarly, our density field $\rho_g(\x)$ has been constructed via a ``galaxies minus randoms'' procedure (Eq.\ (\ref{eq:rhog_def})) which effectively removes modes, since the redshift distribution of the randoms is identical to the galaxies.
(This type of mode removal in a galaxy field is sometimes called the ``radial integral constraint'' \cite{deMattia:2019vdg}.)
To account for this mode removal in $\rho_g(\x)$, we apply the same mean-subtraction step to the surrogate field $S_g(\x)$ that we do to $S_v(\x)$.

\subsection{Photometric redshift errors}
\label{ssec:photoz}

So far, we have assumed spectroscopic redshifts for simplicity.
Extending the velocity reconstruction technique to photometric surveys, which offer broader sky coverage but have less precise redshift information, introduces additional challenges. 
While the statistical power for detecting primordial signatures like PNG increases with access to larger cosmic volumes, measurement sensitivity diminishes on radial scales comparable to or smaller than redshift errors. 
In this section, we describe the pipeline changes needed for a photometric survey.

For each galaxy, let $z_{\rm obs}$, $z_{\rm true}$ denote the observed (photometric) and true redshift, respectively.
Let $\sigma_z$ denote the photo-$z$ error, which we do not assume is the same for all galaxies. (In the DESILS-LRG sample, $\sigma_z$ varies significantly between galaxies -- see Fig.\ \ref{fig:redshift_analysis}.)

The galaxy catalog contains $(z_{\rm obs}, \sigma_z)$ values for each object, but $z_{\rm true}$ is unknown.
For the random catalog, we randomly assign a $(z_{\rm obs}, z_{\rm true}, \sigma_z)$ triple to each object.
To do this, we deconvolve the observed 2-d $(z_{\rm obs}, \sigma_z)$ galaxy catalog, in order to infer the underlying 3-d $(z_{\rm obs}, z_{\rm true}, \sigma_z)$ distribution.

For each object in a photometric catalog, let $\x_{\rm obs}=(\th,z_{\rm obs})$ and $\x_{\rm true}=(\th,z_{\rm true})$ denote the comoving 3-d positions at observed/true redshifts respectively.
The fields $\hv_r(\x)$ and $\rho_g(\x)$ (defined previously in Eqs.\ (\ref{eq:hvr_no_ms}), (\ref{eq:rhog_def})) are defined by placing delta functions at observed (not true) locations:
\begin{align}
\hv_r(\x) &= \sum_b \sum_{i\in b} W_i^v \, \big( \tT(\theta_i) - \tT_b \big) \delta^3(\x-\x_i^{\rm obs})\,, \nn \\
\rho_g(\x) &=
 \bigg( \sum_{i\in \rm gal}\!W_i^g \, \delta^3(\x-\x_i^{\rm obs}) \bigg)
  \!-\!\frac{N_g}{N_r} 
    \bigg( \sum_{j\in \rm rand}\!\!\!W_j^g \, \delta^3(\x-\x_j^{\rm obs}) \bigg)\,.
    \label{eq:photoz_fields}
\end{align}
We also allow the per-object weights $W_i^g$, $W_i^v$ to be functions of two variables ($z_{\rm obs}, \sigma_z)$. (For a spectroscopic catalog, $W_i^g$, and $W_i^v$ would be functions of $z$.)

Our procedure for simulating surrogate fields $S_g(\x)$, $S_v(\x)$ in Eqs.\ (\ref{eq:Sg_sig_def}), (\ref{eq:Sv_sig}) is slightly modified as follows:
\begin{align}
S_g(\x) &\equiv \frac{N_g}{N_r} \sum_{j\in \rm rand} W_j^g \, 
 \delta_G(\x_j^{\rm true}) \, \delta^3(\x-\x_j^{\rm obs}) \,,
 \label{eq:Sg_sig_photo} \\
S_v^{\rm sig}(\x) &\equiv
 \frac{N_g}{N_r} \sum_{j\in\rm rand} 
  W_j^v B(\x_j^{\rm true}) \, v_r(\x_j^{\rm true}) \, \delta^3(\x-\x_j^{\rm obs})\,,
  \label{eq:Sv_sig_photo}
\end{align}
where $B(\x)$ was defined in Eq.~(\ref{eq:B_def}).
That is, delta functions $\delta^3(\x-\x_j)$ are placed at observed locations, whereas random fields $\delta_g$, $v_r$ are evaluated at true locations.
In Appendices \ref{app:surrogates}, \ref{app:hvr_norm} we show that this procedure correctly models the effect of photometric redshift errors on the clustering statistics of $S_g(\x)$ and $S_v(\x)$ respectively.

\section{Data}
\label{sec:data}

\par
\begin{figure*}[t]
    \centering
    \includegraphics[width=1\textwidth]{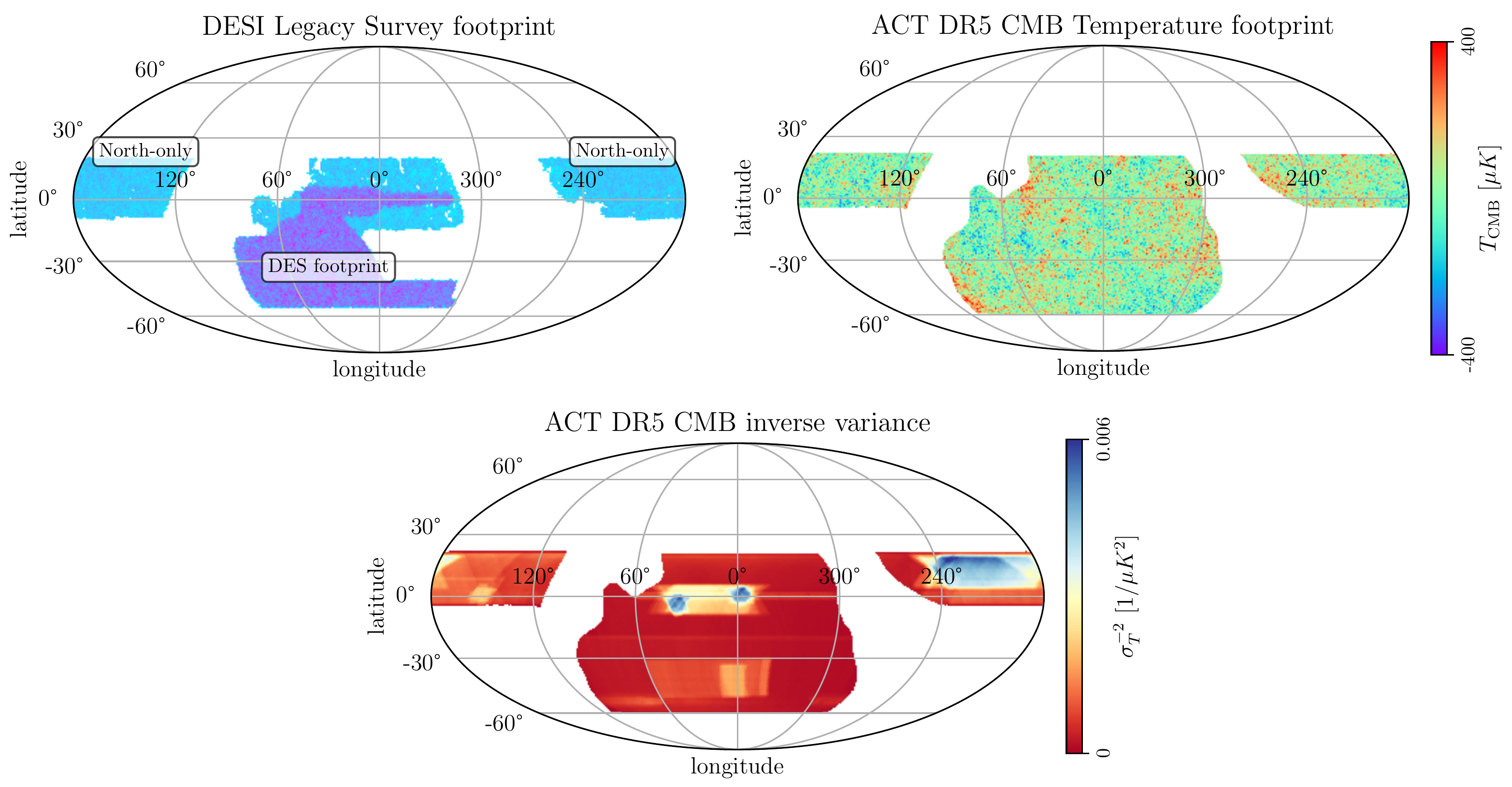}
    \caption{Footprint maps of the galaxy and CMB surveys used in this work. The top left panel shows the DESI-LS footprint; we restrict our analysis to the Galactic northern hemisphere (shown as ``North-only''). This avoids imaging systematics in the South, especially near the DES footprint (purple). The top right panel displays the ACT DR5 150 GHz temperature map, with the color scale indicating measured CMB temperature. The bottom panel shows the inverse variance of the ACT DR5  150 GHz temperature map, where higher fidelity regions appear in blue. 
    Throughout this figure, we mask regions of high Galactic foreground emission, using the Planck $f_{\rm sky}=0.7$ mask.}
    \label{fig:footprint}
\end{figure*}

\noindent
In this section we describe the ACT and DESILS-LRG datasets.
Sky coverage of these datasets is shown in Fig.\ \ref{fig:footprint}.

\subsection{ACT DR5 maps and filtering}

\begin{figure}[t]
    \centering
    \includegraphics[width=1\textwidth,trim={11cm 0 0 0},clip]{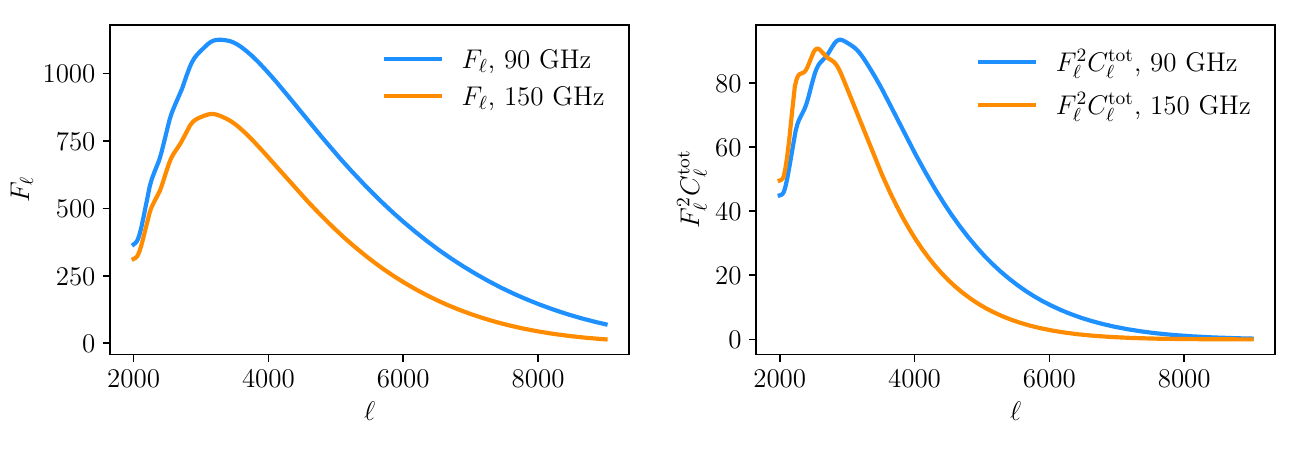}\vspace*{-0.5cm} 
    \caption{{\em Left panel.} CMB filter $F_\ell$ used on the 90 and 150 GHz ACT \texttt{daynight} maps, defined in Eq.\ (\ref{eq:fl_act}). {\em Right panel.} Quantity $(F_\ell^2 C_\ell^{\rm tot})$, which gives the contribution of each multipole $\ell$ to the variance of the filtered CMB. Throughout our filters peak between $\ell\sim2000-4000$ where bulk of our velocity reconstruction signal-to-noise is sourced.} \label{fig:filters_halomodel}
\end{figure}

For CMB maps, we use ACT+\textit{Planck} co-adds at central frequencies of 98 and 150 GHz (former referred to as the `90 GHz map' throughout this paper, as defined in ACT data products). 
We test our analysis on maps that include both nighttime and daytime data (\texttt{daynight} maps) and maps made using data collected exclusively during nighttime observations (\texttt{night} maps). 
The \texttt{daynight} maps are less noisy, but have worse beam systematics due to daytime thermal expansion.\footnote{\hyperlink{https://lambda.gsfc.nasa.gov/product/act/actpol_dr5_coadd_maps_info.html}{See act/actpol\_dr5\_coadd\_maps\_info.html for a more detailed description.}} This makes small-scale ($\ell\gtrsim4000$) CMB temperature less reliable for \texttt{daynight} maps~\citep{Naess:2020wgi}. Since our filters peak around $\ell\sim3000$ (Fig.\ \ref{fig:filters_halomodel}), we find our results are sufficiently insensitive to the beam and we benefit from the inclusion of daytime data and use the \texttt{daynight} maps for our analysis.

Recall from \S\ref{ssec:hvr} that our CMB filter $T_{\rm CMB}(\th) \rightarrow \tT(\th)$ is parameterized by a pixel weight function $W_{\rm CMB}(\th)$ and an $l$-weighting $F_l$.
We decided to match these weightings at 90 and 150 GHz:
\begin{equation}
\label{eq:90_equal_150}
W_{\rm CMB}^{90}(\th) = W_{\rm CMB}^{150}(\th)\,,
  \hspace{1.5cm}
F_l^{150} = \frac{b_l^{90}}{b_l^{150}} F_l^{90}\,,
\end{equation}
where $b_l^\nu$ denotes the CMB beam at frequency $\nu$.
In principle, this matching loses some SNR, but simplifies the analysis by guaranteeing that $b_v^{90} = b_v^{150}$, and that the $(90-150)$ null map is kSZ-free.
This will be useful for the consistency tests in \S\ref{sec:results}.

Following \cite{Lague:2024czc}, we take the pixel weight function to be
\begin{equation}
W_{\rm CMB}(\th) 
 = \begin{cases}
  1 & \mbox{if noise in pixel $\th$ is $\le$ 70 $\mu$K-arcmin, for both 90 and 150 GHz}\,, \\
  0 &\mbox{otherwise}.
 \end{cases}
\end{equation}
We also mask regions of high Galactic foreground emission, by applying the \textit{Planck} $f_{\rm sky}=0.7$ foreground mask.\footnote{\hyperlink{https://irsa.ipac.caltech.edu/data/Planck/release_2/ancillary-data/previews/HFI_Mask_GalPlane-apo0_2048_R2.00/index.html}{HFI\_Mask\_GalPlane-apo0\_2048\_R2.00~(\texttt{GAL070})}} 
We find that masking tSZ clusters does not significantly change our results, provided that the mean-subtraction step in Eq.\ (\ref{eq:hvr_ms}) is used to mitigate CMB foregrounds.

For the CMB $l$-weighting $F_l$, we use Eq.\ (\ref{eq:tT_allsky}) evaluated at 90 GHz:
\begin{equation}
F_l^{90} = \frac{P_{ge}(l/\chi_*)}{P_{gg}(l/\chi_*)} \frac{b_l^{90}}{C_l^{\rm tot}}\,, 
 \hspace{1.5cm}
F_l^{150} = \frac{b_l^{90}}{b_l^{150}} F_l^{90}\,.
 \label{eq:fl_act}
\end{equation}
We set $F_l$ to zero below $l_{\rm min}=2000$ to additionally suppress scales dominated by primary CMB, and above $l_{\rm max}=9000$.
We calculate the power spectra $P_{ge}(k)$, $P_{gg}(k)$ on the RHS using a halo-model calculation (see \S\ref{ssec:fnl} for details).
The CMB filters $F_l^{90}$, $F_l^{150}$ are shown in Fig.\ \ref{fig:filters_halomodel}.

\subsection{DESILS LRG catalogs and processing}

\begin{figure}[t]
    \centering
    \includegraphics[width=1.0\textwidth]{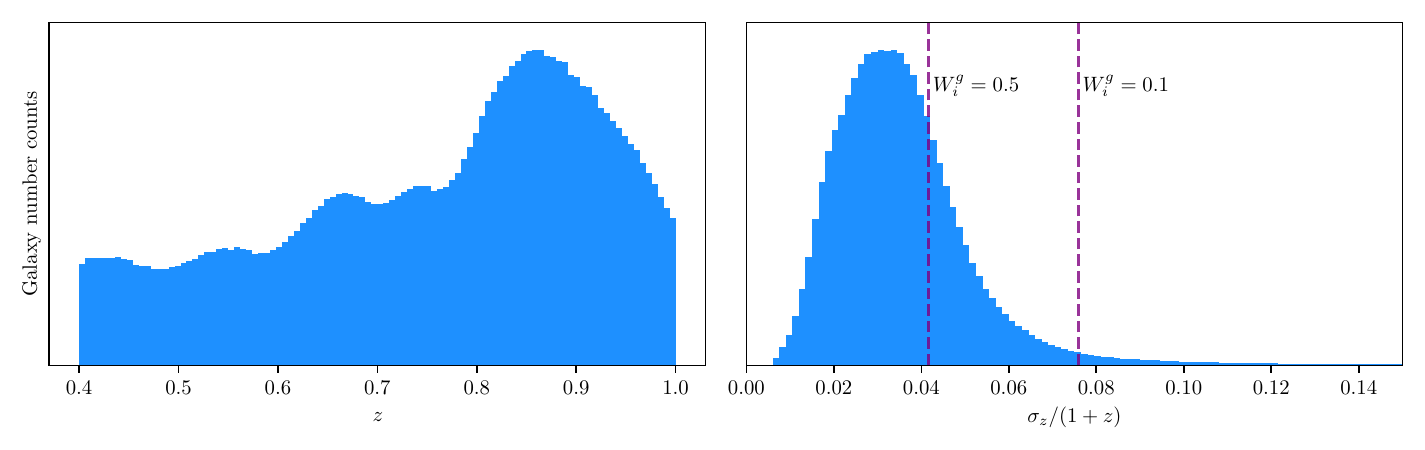}\vspace*{-0.5cm}
    \caption{Histograms of photometric redshifts (left panel) and photo-$z$ errors $\sigma_z/(1+z)$ (right panel) in the DESILS-LRG sample.
    The vertical lines on the right panel represent threshold values of the per-galaxy weight $W_i^g=0.5$~and~0.1, where $W_i^g$ is defined in Eq.\ (\ref{eq:wfkp_photoz}).
    The photo-$z$ error distribution can be seen to show a significant spread in $\sigma_z/(1+z)$ with $W_i^g$ values down-weighting the contribution from galaxies with photo-$z$ errors $\sim0.08$ to our analysis by around 90 percent or more.}
    \label{fig:redshift_analysis}
\end{figure}

For the galaxy survey, we use the LRG sample of DESI Legacy Imaging surveys DR9 as described in Refs.~\citep{DESI:2022gle,Zhou:2023gji}.
We use the ``extended'' LRG sample from \cite{Zhou:2023gji}, with the veto mask and ``quality cuts'' described in \cite{White:2021yvw} (see also the discussion in \S3.3 of \cite{Zhou:2023gji}).
Photometric redshift (photo-$z$) errors in the extended DESI Legacy Survey LRG catalog are typically at the $2–3\%$ level (NMAD $\approx0.02–0.03$), with low outlier fractions, indicating high-quality redshift estimates suitable for large-scale structure and cross-correlation studies.\footnote{The photometric redshift (photo-$z$s) performance of the extended Luminous Red Galaxy (LRG) catalog from the DESI Legacy Imaging Surveys has been evaluated using various machine learning techniques. Notably, a Bayesian Neural Network (BNN) approach, trained on DESI Early Data Release spectroscopic redshifts, achieved a normalized median absolute deviation (NMAD) of approximately 0.026 and an outlier fraction of 0.45\% for LRGs, with a mean uncertainty around 0.0293. Similarly, a Random Forest-based method reported NMAD values near 0.02 for LRGs. These results indicate that the photo-$z$ estimates for LRGs in the extended catalog are of high quality, making them suitable for cosmological analyses such as galaxy clustering and cross-correlations with cosmic microwave background lensing \citep{DESI:2022gle, 2025MNRAS.536.2260Z, Zhou:2023gji}.} The distribution of redshifts and photo-$z$ errors for the DESILS-LRG sample we use in our analysis is shown in Fig.~\ref{fig:redshift_analysis}.

We restrict our analysis to the Galactic northern hemisphere (`North-only' in Fig.~\ref{fig:footprint}) in order to mitigate imaging systematics.
The DESILS combine data from three major optical surveys--DECaLS, BASS, and MzLS--each using different telescopes, instruments, and observing strategies to cover about 14000 deg$^2$ of the sky~\citep{DESI:2016igz, 2019AJ....157..168D}. The heterogeneous mix of instrumentation and observing conditions introduces large-scale systematics, such as variations in depth, photometric calibration, and image quality. These systematics are understood to be largely mitigated through careful analysis~\citep[see e.g.][]{2019AJ....157..168D}. However, we found that the data from the Dark Energy Survey (DES) on the Galactic South patch, which uses the same Dark Energy Camera as DECaLS but employs different photometric processing and calibration methods, lead to visible artifacts such as overdensities or color offsets, and result in systematic effects in our analysis. We choose to expand our analysis into this region in an upcoming work, and for now, only use the Galactic North patch.

In addition to restricting the LRG sample to the Galactic North, we also restrict to redshift range $0.4 \le z_{\rm obs} \le 1.0$, apply the ``quality'' cuts from \S3.3 of \cite{Zhou:2023gji}, and restrict to sky regions where $W_{\rm CMB}(\th) > 0$.
(Note that the $W_{\rm CMB} > 0$ cut includes the Planck Galactic foreground mask.)
The resulting galaxy catalog after these cuts contains 3958201 galaxies with sky area 3442 deg$^2$.
When computing power spectra, we grid these galaxies to a 3-d mesh with pixel size 15 Mpc, and we pad the bounding box by 3000 Mpc to avoid artifacts from periodic boundary conditions.

Recall from Eq.\ (\ref{eq:photoz_fields}) that we define per-galaxy weightings $W_i^g$, $W_i^v$, which are used to construct the $\delta_g$ and $\hv_r$ fields.
In principle, the weightings $W_i^g$, $W_i^v$ are independent, but in our pipeline we choose them to be identical:
\begin{equation}
W_i^g = W_i^v = 
\exp\left( - \frac{\sigma(z_i)^2}{\alpha (1+z_i)^2} \right)
\hspace{1cm} \mbox{where } \alpha=0.0025\,,
\label{eq:wfkp_photoz}
\end{equation}
where $\sigma(z_i)$ is the photometric redshift error of galaxy $i$.
This choice of $W_i$ down-weights objects whose photo-$z$ errors are large compared to the correlation length of the velocity field.

\begin{figure}[t]
    \centering
    \includegraphics[width=0.7\textwidth]{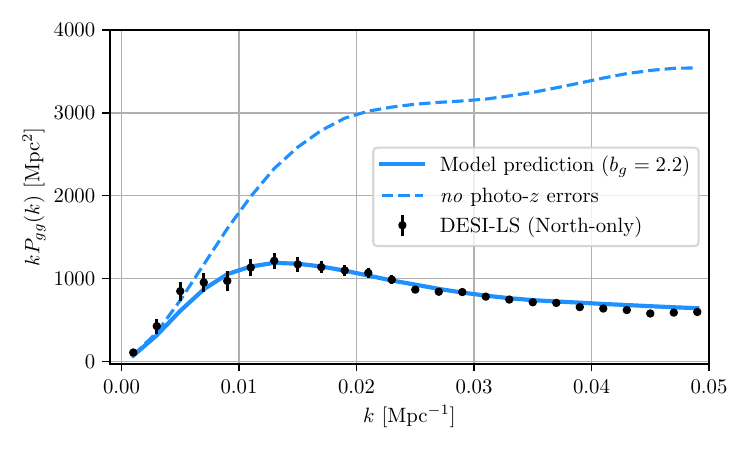}\vspace*{-0.5cm}
    \caption{The galaxy power spectrum $P_{gg}(k)$ from DESILS-LRG (black points), showing good agreement with a linear bias model with Poisson noise (solid curve). This curve was computed by taking the mean power spectrum of simulated surrogate fields $S_g(\x)$ as described in \S\ref{ssec:surr1}. Remarkably, no large-scale systematic power is seen, if the DESILS-LRG catalog is restricted to the Galactic North. Our measurement of $P_{gg}(k)$ is mostly for diagnostic purposes, and is not used to constrain parameters (e.g.\ $\fnl$) in this paper, except for assigning the fiducial value $b_g=2.2$. The dashed curve quantifies the impact of photometric redshift errors on the galaxy power spectrum, by recomputing the model curve assuming zero photo-$z$ error.}
    \label{fig:galaxy_auto}
\end{figure}

\section{Results}
\label{sec:results}

\subsection{Galaxy power spectrum $P_{gg}(k)$}
\label{ssec:pgg}

The galaxy power spectrum $P_{gg}(k)$ is a powerful diagnostic for systematics or modeling issues (such as modelling photo-$z$ errors).
In this paper, we will not infer parameters from $P_{gg}(k)$, except for assigning a fiducial value for the galaxy bias $b_g$, which will be needed to break degeneracies in \S\ref{ssec:fnl}.

In Figure \ref{fig:galaxy_auto}, we show the galaxy power spectrum from the DESILS LRG catalog (points), using the estimator $\hP_{gg}(k)$ defined in Eq.\ (\ref{eq:hpgv_def}).
The solid curve (``model prediction'') is the mean estimated power spectrum of 2000 surrogate fields $S_g(\x)$, constructed as described in \S\ref{sec:high_level_pipeline}.
Surrogate fields were also used to assign error bars in Fig.\ \ref{fig:galaxy_auto}.

The agreement between the data (points) and surrogates (solid line) is a strong test of our surrogate field method.
In particular, it shows that the surrogate field $S_g(\x)$ accurately models photo-$z$ errors, and that the photo-$z$ error estimates $\sigma(z)$ in the DESILS-LRG catalog are not systematically biased.
To quantify this, the dashed line in Fig.\ \ref{fig:galaxy_auto} shows the mean estimated power spectrum of the surrogate fields if photo-$z$ errors are neglected (i.e.\ if we set $\sigma(z)=0$) for all galaxies.
Comparing the solid and dashed lines, we see that the effect of photo-$z$ errors is to suppress power, which makes sense intuitively since photo-$z$ errors can be viewed as radial convolution.\footnote{Note that in Fig.\ \ref{fig:galaxy_auto}, we do not attempt to ``deconvolve'' the power spectrum estimator, i.e.\ correct for the $P_{gg}$ bias due to photo-$z$ errors. Instead, we leave the estimator $\smash{\hP_{gg}(k)}$ in the simple form in Eq.\ (\ref{eq:hpgv_def}), and account for photo-$z$ errors in our model prediction by including them in the surrogate simulations (solid curve). This can be interpreted as ``convolving'' the model, rather than ``deconvolving'' the estimator (see discussion at the end of \S\ref{ssec:pgv}). A similar comment applies to measurements of the galaxy-velocity power spectrum $P_{gv}(k)$ to be presented shortly (Fig.\ \ref{fig:ksz_velocity}).}

One striking feature of Fig.\ \ref{fig:galaxy_auto} is the absence of systematic power on large scales.
Previous studies have found that mitigating imaging systematics in DESILS is challenging and requires sophisticated methods.
For example, in \cite{Rezaie:2023lvi} a nonlinear, neural network based method was used to mitigate systematic large-scale power.
We find that imaging systematics seem to be a non-issue if the DESILS-LRG catalog is restricted to the Galactic North (``North-only'' in Fig.\ \ref{fig:footprint}), where the DESILS survey is more uniform.
(We tried including data from some or all of the South, but found significant imaging systematics in $P_{gg}$ and other diagnostic plots.)
In this paper, we decided to use only the North data for simplicity. 
In future work, we hope to include South data, perhaps by adopting some of the sophisticated methods referenced above.

\subsection{Cross power spectrum $P_{gv}(k)$}

\begin{figure}[t!]
    \centering
    \includegraphics[width=0.7\textwidth]{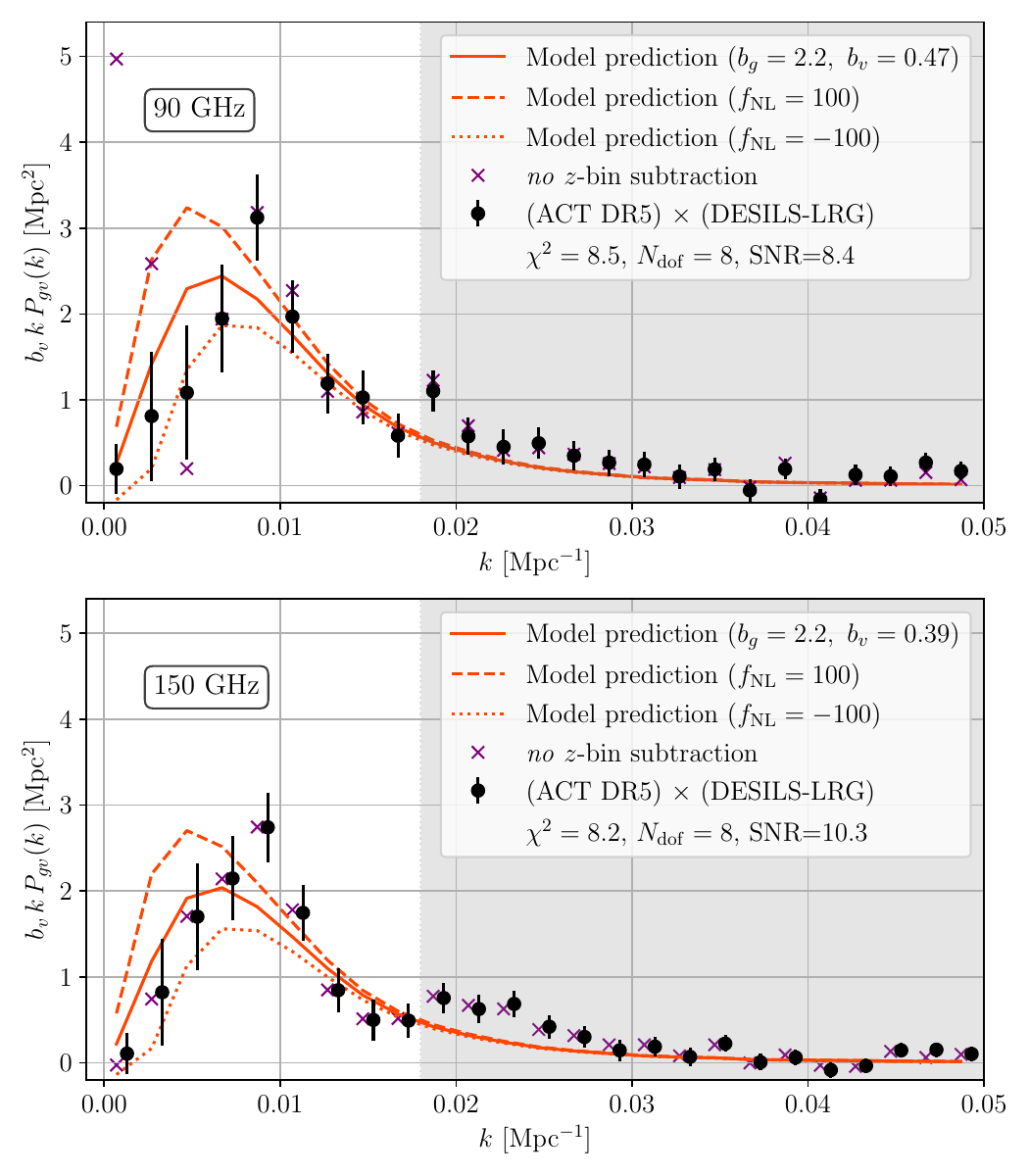}\vspace*{-0.4cm}
    \caption{Galaxy-velocity cross-power spectra $P_{gv}(k)$ using the 90 GHz (top) and 150 GHz (bottom) CMB maps from ACT DR5 and the DESI Legacy Survey (black points with error bars).
    Solid red lines show model predictions with $b_g=2.2$, $\fnl=0$, and $b_v = \{0.47,0.39\}$ for $\{90,150\}$ GHz.
    This model is a good fit ($\chi^2 = 8.5$ or 8.2 with $N_{\rm dof}=8$) over the range of scales used in this paper ($k \lesssim 0.018$ Mpc$^{-1}$, delimited by the grey band).
    The dashed and dotted lines correspond to our model prediction with $f_{\rm NL}=\pm 100$.
    The signal-to-noise ratio (SNR) is $\{8.4, 10.3\}$ at $\{90,150\}$ GHz.
    The purple cross markers show the estimated $P_{gv}(k)$ if we omit the $\hat v_r$ mean-subtraction step in Eq.~\eqref{eq:hvr_ms}.
    The large difference between the black points and purple crosses suggests significant foreground contamination at 90 GHz, which is suppressed by the mean-subtraction step.}
    \label{fig:ksz_velocity}
\end{figure}

In Figure \ref{fig:ksz_velocity}, we show a central result of the paper: the cross power spectrum $P_{gv}(k)$ between the DESILS-LRG galaxy field $\rho_g$, and the kSZ velocity reconstruction $\hat v_r$ obtained from ACT$\times$DESILS.
Throughout the paper, we use the first nine $k$-bins (corresponding to $k \lesssim$ 0.018 Mpc$^{-1}$) for consistency checks and model fitting, including the $\fnl$ constraints in \S\ref{ssec:fnl}.

Cross power spectra were estimated using the estimator $\hP_{gv}(k)$ defined in Eq.\ (\ref{eq:hpgv_def}).
Note that this estimator estimates the quantity $(b_v P_{gv}(k))$, with an overall factor of $b_v$.
It is not an unbiased estimator, e.g.\ $\smash{\hP_{gv}(k)}$ is suppressed at high $k$ due to photometric redshift errors (just like $\smash{\hP_{gg}(k)}$ in Fig.\ \ref{fig:galaxy_auto}).
This is not a problem for parameter constraints, since we self-consistently include this high-$k$ suppression in our surrogate simulations.
As in the previous section, solid curves in Fig.\ \ref{fig:ksz_velocity} are model predictions for $P_{gv}(k)$, computed by taking the mean estimated cross power spectrum of surrogate fields $S_g(\x)$, $S_v(\x)$ as described in \S\ref{sec:high_level_pipeline}.

Our model for $P_{gv}(k)$ has three parameters $(b_g, b_v, \fnl)$.
In Fig.\ \ref{fig:ksz_velocity}, we have taken $b_g=2.2$, $\fnl=0$, and best-fit values $b_v = \{0.47, 0.39\}$ at $\{90,150\}$ GHz respectively.
These low values of $b_v$ (compared to $b_v=1$) are interpreted in \S\ref{ssec:fnl}.

\begin{figure}[t!]
    \centering
    \includegraphics[width=0.7\textwidth]
    {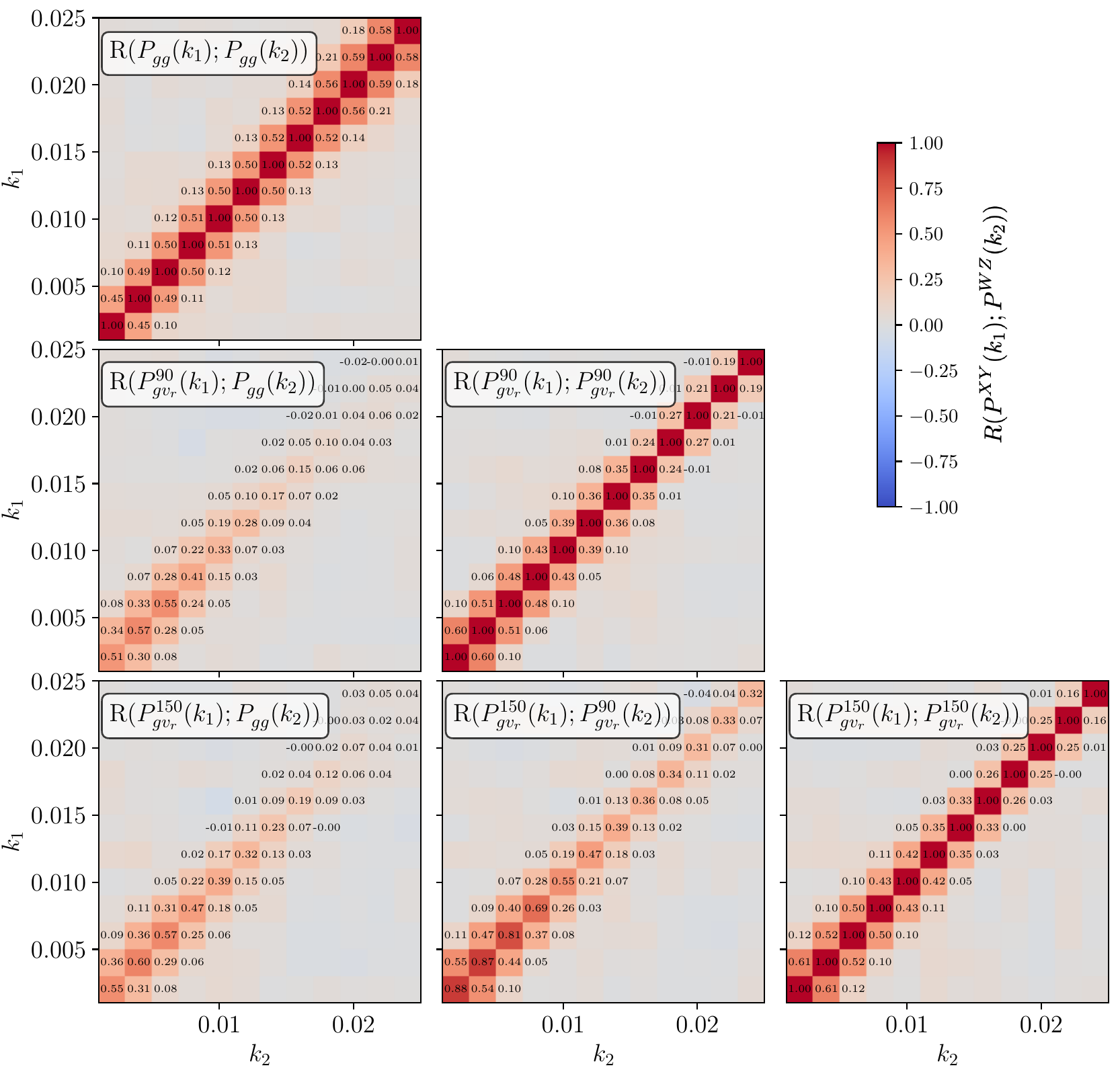}\vspace*{-0.36cm}
  \caption{Visualizing the band-power correlations of $P_{gg}(k_1)$ and $P_{gv}(k_2)$ from 90~and~150\,GHz ACT DR5 \texttt{daynight} CMB maps, calculated using our surrogate simulations. The R matrix is defined in Eq.~\eqref{eq:rmatrix}. Significant correlations are seen between adjacent $k$-bins, between $P_{gv}^{90}$ and $P_{gv}^{150}$, and between all fields at very low $k$.}
    \label{fig:covariance}
\end{figure}

To assess whether this model is a good fit, we compute the $\chi^2$ statistic separately for 90 and 150 GHz:
\begin{equation}
\chi^2 = (d-m)^T C^{-1} (d-m)\,,
\label{eq:chi2_def}
\end{equation}
where $d$ is a nine-component vector representing the nine ``data'' bandpowers in Fig.\ \ref{fig:ksz_velocity}, $m$ is the best-fit model, and $C$ is the 9-by-9 covariance matrix evaluated at the best-fit model.
The off-diagonal correlations in $C$ are shown visually in Fig.\ \ref{fig:covariance}.
We get $\chi^2=\{ 8.5, 8.2 \}$ at \{90,150\} GHz, with eight degrees of freedom (since there are nine bandpowers, and one fitted parameter $b_v$), which is an excellent fit ($p$-value $\{0.38, 0.41\}$).
These excellent fits are a strong check that our surrogate simulations are correctly estimating error bars on $P_{gv}(k)$.

The signal-to-noise (SNR) values shown in Fig.\ \ref{fig:ksz_velocity} are defined by:
\begin{equation}
\mbox{SNR} = \frac{d^T C_0^{-1} m}{[m^T C_0^{-1} m]^{1/2}}\,,
\label{eq:snr}
\end{equation}
where $d,m$ have the same meaning as in Eq.\ (\ref{eq:chi2_def}), and $C_0$ is the 9-by-9 bandpower covariance in a model with $b_v=0$ (i.e.\ such that $\rho_g$ and $\hv_r$ are uncorrelated).

Next, we explain the meaning of the ``no $z$-bin subtraction'' points (crosses) in Fig. \ref{fig:ksz_velocity}.
Recall that in \S\ref{ssec:hvr} (Eqs.\ (\ref{eq:hvr_no_ms}), (\ref{eq:hvr_ms})), we defined $\hat v_r$ with a ``mean subtraction'' step, in order to mitigate CMB foregrounds.
The crosses in Fig.\ \ref{fig:ksz_velocity} show the power spectrum $P_{gv}(k)$ if this mean subtraction step is omitted.
We see that the 150 GHz measurements are barely affected, while the 90 GHz measurements change dramatically.
(We show in \S\ref{ssec:null_tests} that the 90 and 150 GHz measurements are statistically consistent, if both are mean-subtracted.)

Our surrogate-based pipeline estimates the covariance between $k$-bins, and between power spectra of different types (either $P_{gg}(k)$, $P_{gv}^{90}(k)$, or $P_{gv}^{150}(k)$).
In Fig.\ \ref{fig:covariance}, we show correlation coefficients, defined by:
\begin{equation}\label{eq:rmatrix}
    R(P^{XY}(k_1);P^{WZ}(k_2))=\frac{{\rm Cov}(P^{XY}(k_1);P^{WZ}(k_2))}{\sqrt{{\rm Var}(P^{XY}(k_1))\, {\rm Var}(P^{WZ}(k_2))}}\,,
\end{equation}
where $\{P^{XY}(k),P^{WZ}(k)\}\in\{P_{gg}(k),P_{gv}^{90}(k),P_{gv}^{150}(k)\}$.

\subsection{Constraints on $f_{\rm NL}$ and $b_v$}
\label{ssec:fnl}

In this section, we use the $P_{gv}(k)$ measurements from Fig.\ \ref{fig:ksz_velocity}, and the bandpower covariance in Fig.\ \ref{fig:covariance}, to constrain $\fnl$ and the kSZ velocity bias $b_v$.
Our basic tool is Monte Carlo Markov chain (MCMC) sampling of the Gaussian bandpower likelihood ${\mathcal L}$, defined by:
\begin{equation}
\log\mathcal{L}(\theta) = 
 -\frac{1}{2} \log\det\mathcal{C(\theta)} 
 -\frac{1}{2} \big( \boldsymbol{t}(\theta)-\boldsymbol{d} \big)^T \mathcal{C(\theta)}^{-1}
 \big( \boldsymbol{t}(\theta)-\boldsymbol{d} \big)
 + \textrm{const}\,,
 \label{eq:likelihood}
\end{equation}
Here, $\boldsymbol{d}$ is a ``data'' vector containing $P_{gv}(k)$ bandpowers, with either nine components (if we are analyzing 90 or 150 GHz separately), or 18 components (if we are combining the 90 and 150 GHz bandpowers into a joint likelihood).
The vector $\boldsymbol{t}(\theta)$ is a ``theory'' prediction for $P_{gv}(k)$ given parameters $\theta \in \{b_v,f_{\rm NL}\}$, which we model using surrogate fields as described in \S\ref{sec:high_level_pipeline}.

The matrix ${\mathcal C}(\theta)$ in Eq.\ (\ref{eq:likelihood}) is the $P_{gv}$ bandpower covariance, which we also model using surrogate fields.
Note that our surrogate field machinery computes all entries of ${\mathcal C}$, including correlations between $P_{gv}^{90}(k)$ and $P_{gv}^{150}(k)$ in the case of a joint likelihood (Fig.\ \ref{fig:covariance}), and including dependence on parameters $\theta = \{ \fnl, b_v \}$.
(Incorporating the full $\fnl,b_v$ dependence of ${\mathcal C}$ turned out not to be very important, but it is important to evaluate ${\mathcal C}$ at parameter values that are reasonable fits to the data.)

We will fix maximum wavenumber $k_{\rm max} = 0.018\,{\rm Mpc}^{-1}$ and fiducial galaxy bias $b_g=2.2$ (from Fig.\ \ref{fig:galaxy_auto}) throughout.
In future work, we plan to implement a joint $P_{gg}+P_{gv}$ analysis to properly estimate the statistical error on $b_g$.
In this paper, where we constrain parameters from $P_{gv}$ only, we assume a fiducial value of $b_g$ in order to break degeneracies (otherwise, $P_{gv}(k)$ would only constrain the parameter combinations $(b_g b_v)$ and $(b_g-1) b_v \fnl$).

The total SNR (defined previously in Eq.\ (\ref{eq:snr}) is given by:
\begin{equation}
\mbox{SNR} = \begin{cases}
 8.4 & \mbox{90 GHz only}, \\
 10.3 & \mbox{150 GHz only}, \\
 11.7 & \mbox{90+150 GHz joint analysis}.
\end{cases}
\end{equation}
This detection SNR is similar to other kSZ measurements based on stacking, e.g.\ \cite{Hadzhiyska:2024qsl}. 

\begin{figure}[t]
    \centering
    \includegraphics[width=0.58\textwidth]
    {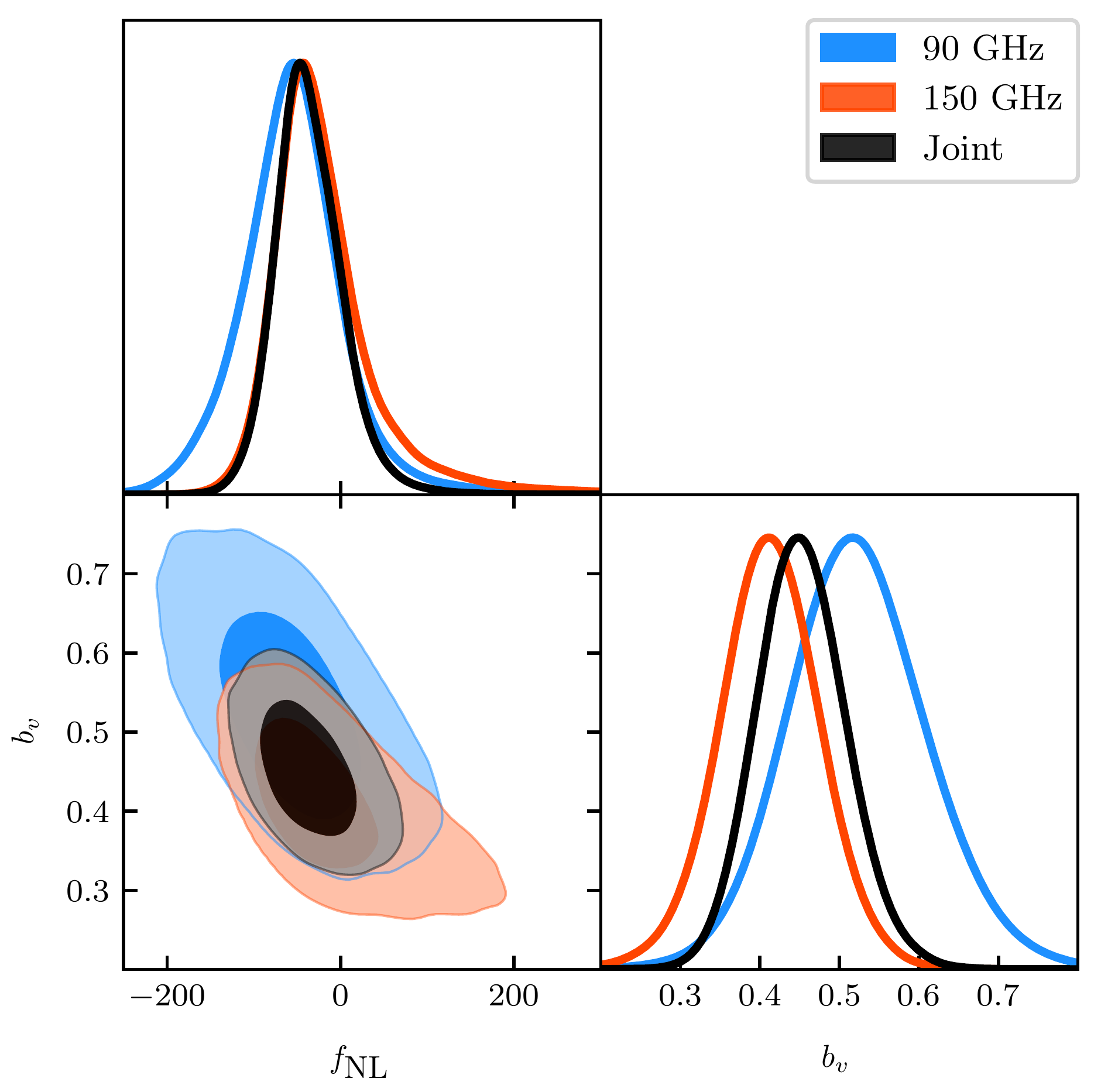}
    \vspace*{-0.34cm}
    \caption{Results from our Markov chain Monte Carlo (MCMC) runs using DESI-LS `North-only' patch (see Fig.~\ref{fig:footprint}) and ACT DR5 \texttt{daynight} maps. The blue and orange contours correspond to velocity reconstruction using 90~and~150\,GHz maps respectively. The black contour corresponds to our joint 90+150 analysis, where we include both 90~and~150\,GHz bandpowers and their covariance. The 1d-contours correspond to marginalized a-posteriori distributions of parameters $f_{\rm NL}$ and $b_v$. The results of these runs are shown in Table~\ref{tab:constraints_table}. We find $f_{\rm NL}=-39^{+40}_{-33}$, consistent with zero and around a factor five more stringent than earlier results using 3-d velocity reconstruction. 
    We find a low bias, $b_v=0.45^{+0.06}_{-0.05}$, which aligns with earlier findings of e.g.\ Ref.~\citep{Hadzhiyska:2024qsl}, with higher feedback driving the true electron-galaxy correlation $P_{ge}^{\rm true}(k)$ to be lower than to our fiducial $P_{ge}^{\rm fid}(k)$.}
    \label{fig:ksz_fnl}
\end{figure}

We use the \texttt{emcee} software~\citep{2013PASP..125..306F} to run MCMCs in the parameter space $\{\fnl,b_v\}$.
Posterior likelihoods are plotted in Fig.\ \ref{fig:ksz_fnl}.
The $1\sigma$ marginalized constraints on $\fnl$ are given by:
\begin{equation}
\fnl = \begin{cases}
 -55.4^{+52.9}_{-54.4} & \mbox{90 GHz only}, \\
 -30.2^{+49.9}_{-39.5} & \mbox{150 GHz only}, \\
 -39.3^{+40.2}_{-33.4} & \mbox{90+150 GHz joint analysis}.
\end{cases}
\label{eq:bottom_line_fnl}
\end{equation}
We find the large volume and galaxy number count of DESI-LS result in around-a-factor-five higher-fidelity constraints on $f_{\rm NL}$ compared to previous constraints, \textit{e.g.}~in~Ref.~\citep{Lague:2024czc}, which used the SDSS spectroscopic galaxy catalog.
The 1$\sigma$ marginalized constraints on $b_v$ from our MCMCs are given by:
\begin{equation}
b_v = \begin{cases}
 0.52^{+0.09}_{-0.08} & \mbox{90 GHz only}, \\
 0.41^{+0.06}_{-0.06} & \mbox{150 GHz only}, \\
 0.45^{+0.06}_{-0.05} & \mbox{90+150 GHz joint analysis}.
\end{cases}
\label{eq:bottom_line_bv}
\end{equation}
Recall that the value of $b_v$ can be interpreted (Eq.\ (\ref{eq:bv_def})) as the ratio between the true galaxy-electron power spectrum $P_{ge}^{\rm true}(k)$ and the fiducial $P_{ge}^{\rm fid}(k)$.
In order to interpret our measurement (\ref{eq:bottom_line_bv}) of $b_v$, we explain in more detail how our $P_{ge}^{\rm fid}(k)$ is computed.
We use the halo model prescription in the \texttt{hmvec} code\footnote{https://github.com/simonsobs/hmvec} \cite{Smith:2018bpn}, and compute $P_{ge}(k,z)$ at the central redshift $z_*=0.7$ of DESILS-LRG.
In the halo model calculation, we use the Battaglia gas profile~\cite{Battaglia:2016xbi}, also referred to as the `AGN' profile in the literature.
We use an HOD defined by abundance matching to the mean galaxy density $n_g = 3 \times 10^{-4}$ Mpc$^{-3}$ of our LRG catalog after quality cuts.
We find that $P_{ge}^{\rm fid}$ is not very sensitive to the choice of $n_g$ (the derivative $d(\log P_{ge}^{\rm fid})/d(\log n_g)$ is $\approx 0.3$).

Therefore, the value (\ref{eq:bottom_line_bv}) of $b_v$ less than unity shows that the gas profiles from \cite{Battaglia:2016xbi} overestimate $P_{ge}^{\rm true}(k)$ on scales where we are sensitive ($2000 \lesssim \ell \lesssim 4000$). 
This finding aligns with earlier results \cite{Hadzhiyska:2024qsl, Hadzhiyska:2024ecq, RiedGuachalla:2025byu} that show evidence of enhanced feedback.
In future work, it would be interesting to compare with a wider variety of predictions for $P_{ge}(k)$, for example from the Illustris~\cite{Nelson:2018uso} or Flamingo \cite{Schaye:2023jqv} simulations.

\subsection{Null tests}
\label{ssec:null_tests}

\begin{figure}
    \centering
    \includegraphics[width=0.95\textwidth]
    {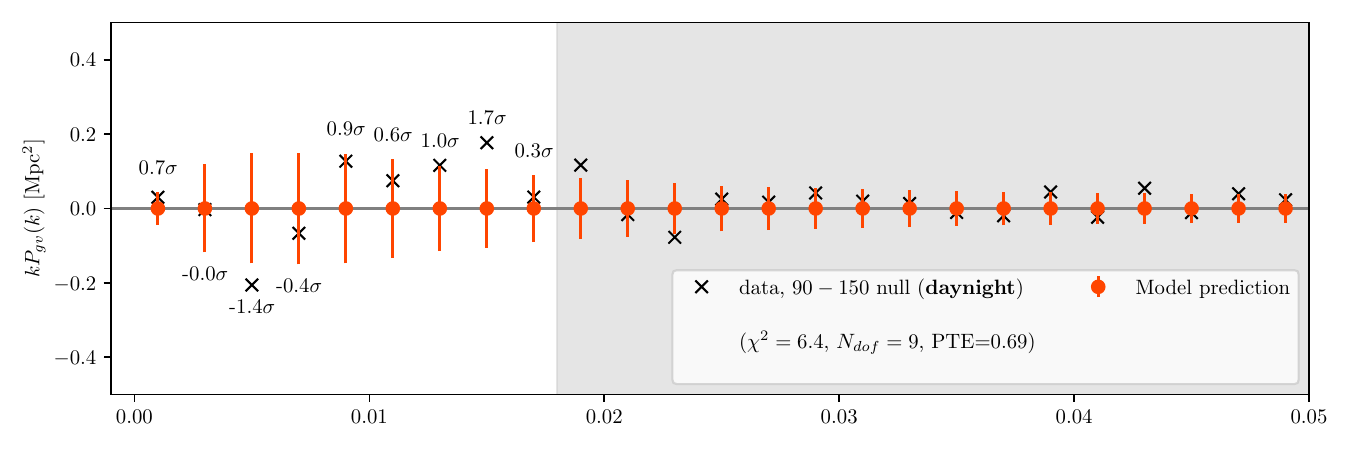}\vspace*{-0.36cm}
    \includegraphics[width=0.95\textwidth]
    {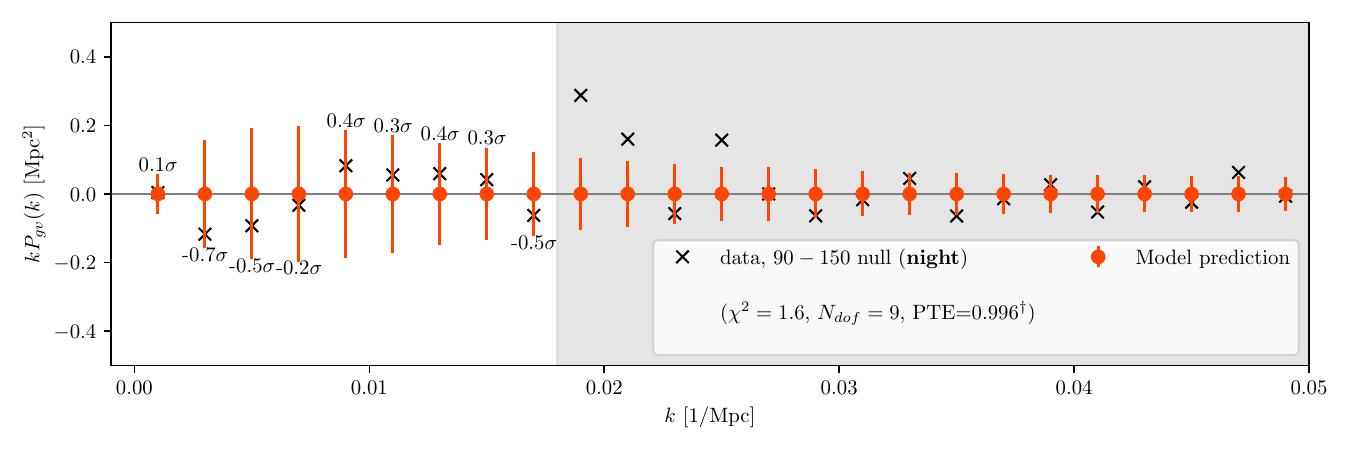}    
    \vspace*{-0.3cm}
    \caption{Null tests of the galaxy–velocity spectrum difference between 90~and~150\,GHz using the DESI‐LS `North‐only' region (see Fig.~\ref{fig:footprint}) with ACT‐DR5 \texttt{daynight} and \texttt{night} maps (top and bottom panels, respectively). The 90~and~150\,GHz maps are filtered so that their difference is kSZ‐free (Eq.~\eqref{eq:90_equal_150}). We define 
    $P_{gv}^{\rm null}(k)\equiv P_{gv}^{90}(k)-P_{gv}^{150}(k)\,,$ and show the measurements with black crosses. Orange points and error bars indicate the surrogate‐field model prediction (i.e., zero). Taking data points until $k\approx0.018{\rm Mpc}^{-1}$, we find both null spectra are consistent with zero, with PTEs of 0.69 (\texttt{daynight}) and 0.996 (\texttt{night}). Note that the PTE of the \texttt{night} map reduces to 0.39 if we include the next data point at $k=0.019$ Mpc$^{-1}$.}
    \label{fig:null_test}
    \vspace*{-0.3cm}
\end{figure}

\begin{table}
\begin{center}
\begin{tabular}{|| c | l | l | c | c ||} 
 \hline  & & & & \\ [-2ex]
 ACT DR5 CMB & Marg. post. $f_{\rm NL}$  & Marg. post.  $b_v$  & SNR & $\chi^2$ \\ [1ex] \hline\hline & & & & \\ [-1.5ex]
 Joint (\texttt{daynight}) & $-39.3^{+40.2\,(+91.9)}_{-33.4\,(-67.8)}$ & $0.45^{+0.06\,(+0.12)}_{-0.05\,(-0.11)}$ & 11.7 & 9.7 \\ [1ex] 
 Joint (\texttt{night}) & $-64.6^{+42.4\,(+82.7)}_{-50.0\,(-92.6)}$ & $0.43^{+0.07\,(+0.14)}_{-0.06\,(-0.12)}$ & 8.6 & 4.8 \\ [1ex] 
 \hline
 150 (\texttt{daynight}) & $-30.2^{+59.9\,(+202.2)}_{-39.5\,(-78.1)}$ & $0.41^{+0.06\,(+0.13)}_{-0.06\,(-0.13)}$ & 10.3 & 8.6  \\ [1ex] 
 150 (\texttt{night}) &  $-54.9^{+51.6\,(+121.7)}_{-47.6\,(-97.1)}$ & $0.41^{+0.08\,(+0.16)}_{-0.07\,(-0.14)}$ & 7.6 &  2.0  \\ [1ex] 
 \hline
 90 (\texttt{daynight}) & $-55.4^{+52.9\,(+143.9)}_{-54.4\,(-117.5)}$ & $0.52^{+0.09\,(+0.19)}_{-0.08\,(-0.17)}$ & 8.4 & 9.0 \\ [1ex] 
 90 (\texttt{night}) & $-85.4^{+62.9\,(+138.1)}_{-54.6\,(-107.3)}$ &  $0.45^{+0.10\,(+0.20)}_{-0.09\,(-0.18)}$ & 5.8 & 4.1 \\ [0.1ex]
 \hline
 \end{tabular}
 \end{center}
 \caption{Measured constraints on squeezed 
 primordial non-Gaussianity amplitude $f_{\rm NL}$ and the optical depth bias $b_v$, along with SNR and chi-square ($N_{\rm dof}=8$) from velocity-galaxy cross-correlation; shown for velocity reconstruction using \texttt{daynight} and \texttt{night} maps. The {\tt daynight} and {\tt night} results are consistent.}
 \label{tab:constraints_table}
\end{table}

\begin{table}
\begin{center}
\begin{tabular}{|| c | c | c | c ||} 
 \hline
 CMB  & Test Description & PTE (\texttt{daynight}) & PTE (\texttt{night}) \\ [0.5ex] 
 \hline\hline
 $90$ &  DESI-LS \& mock CMB & 0.67 & 0.85 \\ 
 \hline
 $150$ &  DESI-LS \& mock CMB & 0.35 & 0.26 \\ 
 \hline
 $150-90$ &  DESI-LS \& mock CMB & 0.53 & 0.22 \\ 
 \hline
 Joint &  DESI-LS \& mock CMB & 0.59 & 0.91 \\ 
 \hline
 \end{tabular}
 \end{center}
 \vspace*{-0.5cm}
 \caption{The $p$-value (PTE) results from our null tests. We find all of our null tests satisfy the null condition (${\rm PTE}>0.05$).}
 \label{tab:null_test_table}
\end{table}

Finally, we discuss the various null tests we performed to validate our analysis. 

Most importantly, in Fig. \ref{fig:null_test} we test for CMB foreground contamination by comparing null power spectra $(P_{gv}^{90}(k) - P_{gv}^{150}(k))$ with zero.
Since the kSZ signal has a black-body frequency dependence, the `null' power spectra have zero contribution from kSZ, once we equalize the beams of the 90~and~150\,GHz maps in our analysis (see Eq.~\eqref{eq:90_equal_150}). We display the `null' power spectra results in Fig.~\ref{fig:null_test}. 
For both of the \texttt{daynight} and \texttt{night} maps, taking the first 9 $k$-bins up to $k_{\rm max} = 0.018$ Mpc$^{-1}$ as described above, we find the null spectra agree with zero signal at high significance with PTEs $\approx0.69$~and~0.996, respectively. 
This latter value (PTE=0.996) is close to 1, which means that the null power spectrum is closer to zero than statistically expected (i.e. ``too good'').
We interpret this as a statistical fluke, since it happens for the {\tt night} null map but not {\tt daynight}, and if we include the next data point (at $k=0.019$ Mpc$^{-1}$), the null spectra PTE of the \texttt{night} map reduces to 0.39. 

Next, in Table \ref{tab:constraints_table} we test for consistency between ACT {\tt daynight} and {\tt night} ACT maps.
We rerun our MCMC pipeline from \S\ref{ssec:fnl} on {\tt night} maps, and verify that the results are consistent with the {\tt daynight} maps (also shown in the table).

Finally, in Table \ref{tab:null_test_table}, we rerun the pipeline with ``mock'' CMB maps obtained by rotating the \texttt{daynight} and \texttt{night} ACT CMB maps by 4 degrees north in declination.
The corresponding joint region matches our baseline and we find the CMB-galaxy correlation to vanish while the galaxy autopower spectrum remains the same. We analyze the 90~and~150\,GHz maps, as well as the null galaxy-reconstructed--velocity power spectra and the joint analysis of both maps, as previously described. We confirm that for all our null tests, setting $b_v=0$, the null condition (that is, PTE$>0.05$) is satisfied to high significance. See Table~\ref{tab:null_test_table} for the list of our PTEs.

\section{Discussion}

In this paper, we have presented the first 3-d kSZ velocity reconstruction using a photometric galaxy sample, using (ACT DR5) $\times$ (DESILS LRGs), restricted to the northern galactic hemisphere in order to mitigate imaging systematics.
Our analysis represents the first of its kind, leveraging the full three-dimensional galaxy density field derived from photometric redshift measurements. This approach enables novel synergies between cosmic microwave background (CMB) observations and large-scale structure surveys, facilitating unprecedented cross-correlations between disparate cosmological probes. 

We find SNR $11.7\sigma$, matching other kSZ-based analyses on similar datasets \cite{Hadzhiyska:2024qsl}. 
Our constraints on $f_{\rm NL}$ are the most stringent to date ($f_{\rm NL}\!=\!-39^{+40}_{-33}$) from the emerging velocity tomography program.
We also find that the overall amplitude of the galaxy-electron power spectrum is lower than a halo model prediction ($b_v = 0.45^{+0.06}_{-0.05}$).
Our measurement of $P_{gv}(k)$ passes many goodness-of-fit and null tests, including consistency (within statistical errors) between 90 GHz and 150 GHz (\S\ref{ssec:fnl}, \S\ref{ssec:null_tests}).

In future work, we plan to extend this analysis in several directions as follows:

\begin{itemize}

\item
In this paper, we considered the galaxy-velocity power spectrum $P_{gv}(k)$, in order to report parameter constraints from kSZ alone, without including information from galaxy clustering.
Future applications of this methodology (e.g.\ in LSST) will combine the $P_{gg}$, $P_{gv}$, and $P_{vv}$ power spectra to cancel large-scale cosmic variance, thereby enhancing statistical precision. This technique is expected to surpass the constraints on PNG obtained from $P_{gg}$ alone in the near future \cite{Munchmeyer:2018eey}.

\item
We did not make a systematic effort to construct optimal estimators or maximize SNR.
(In particular, we chose to throw out the Galactic southern hemisphere, in order to reduce imaging systematics and pass null tests.)
It would be interesting to construct optimal $(S+N)^{-1}$ weighted estimators and compare, especially in the high SNR regime where sample variance cancellation is important.

\item
We introduced surrogate fields as a technical tool, in order to model the effect of the survey window function on the largest scales, and to estimate error bars.
In work in progress, we are developing new tools to model the large-scale window function, and to make non-Gaussian mock catalogs in very large volumes.
In particular, mock catalogs with $f_{NL}\ne 0$ will be valuable for Monte Carlo pipeline validation, and mocks with simulated CMB foregrounds will be valuable for exploring the impact of foregrounds.

\item
We have neglected RSDs, since we expect their effect to be small for a photometric survey, but properly accounting for RSDs will be important for future analyses involving spectroscopic surveys (e.g.\ DESI).

\end{itemize}

Our measurement introduces a powerful framework for testing fundamental physics, including early-universe models and inflationary dynamics. More broadly, this work establishes a novel observational paradigm that integrates CMB secondary anisotropies with the statistical power of photometric large-scale structure surveys, paving the way for transformative advancements in precision cosmology.

\section*{Acknowledgments}
We thank Edmond Chaussidon, Boryana Hadzhiyska, Matthew Johnson, Yurii Kvasiuk, Anderson Lai, Alex Lagu\"e, Dustin Lang, Mathew Madhavacheril, Moritz M\"unchmeyer, Noah Sailer, and Rongpu Zhou for very useful discussions. 
SCH was supported by the P.~J.~E.~Peebles Fellowship at Perimeter Institute. KMS was supported by an NSERC Discovery Grant, by the Daniel Family Foundation, and by the Centre for the Universe at Perimeter Institute. Research at Perimeter Institute is supported by the Government of Canada through Industry Canada and by the Province of Ontario through the Ministry of Research \& Innovation.
SF is supported by Lawrence Berkeley National Laboratory and the Director, Office of Science, Office of High Energy Physics of the U.S. Department of Energy under Contract No.\ DE-AC02-05CH11231.

\bibliographystyle{apsrev4-1}
\bibliography{long_paper}

\appendix

\section{Power spectrum normalization}
\label{app:power_spectrum_normalization}

The purpose of this appendix is to explain how the power spectrum normalizations $\N_{gg}$, $\N_{gv}$ defined in Eq.\ (\ref{eq:hpgv_def}) are computed.
We emphasize that our constraints on $f_{NL}$ and $b_v$ do not depend on the values of $\N_{gg}$, $\N_{gv}$.
Parameter constraints are obtained by comparing values of $\hP_{gv}(k)$ on data and surrogate simulations, and are independent of the overall normalization of $\hP_{gv}(k)$.
The normalizations $\N_{gg}$, $\N_{gv}$ are only used so that power spectrum plots (Figs \ref{fig:galaxy_auto}, \ref{fig:ksz_velocity}) will have a sensible normalization.

First, some general comments on power spectrum normalization in a finite survey volume.
Let $\phi(\x)$, $\phi'(\x)$ be random fields, and suppose we are interested in the cross power spectrum $P_{\phi\phi'}(k)$.
The simplest power spectrum estimator $\hP_{\phi\phi'}(k)$ which makes sense in a finite survey volume is defined by:
\begin{equation}
\hP_{\phi\phi'}(k) \equiv \int \frac{d\Omega_k}{4\pi} \,
 \tphi(\k)^* \tphi'(\k)
\hspace{1cm}
\tphi(\k) \equiv \int d^3\x \, W(\x) \phi(\x) 
\hspace{1cm}
\tphi'(\k) \equiv \int d^3\x \, W'(\x) \phi'(\x) 
\end{equation}
where the (non-random) weight functions $W(\x)$, $W'(\x)$ represent the survey geometry (which can be different for the two fields).

How is the estimated power spectrum $\hP_{\phi\phi'}(k)$ related to the true power spectrum $P_{\phi\phi'}(k)$?
A short calculation shows:
\begin{equation}
\big\langle \hP_{\phi\phi'}(k) \big\rangle
 = \int \frac{d\Omega_k}{4\pi} \int \frac{d^3\k'}{(2\pi)^3} \,
   \big[ W(\k-\k') W'(\k-\k')^* \big] \,
   P_{\phi\phi'}(k') \label{eq:k_mixing}
\end{equation}
For slowly varying window functions, the quantity $\big[ W(\k-\k') W'(\k-\k')^* \big]$ will be a narrowly peaked function of its argument $(\k-\k')$, and we can make the approximation:
\begin{equation}
\big\langle \hP_{\phi\phi'}(k) \big\rangle
 \approx \N_{WW'} P_{\phi\phi}(k)
\hspace{1.5cm} \mbox{where }
 \N_{WW'} \equiv \int \frac{d^3\K}{(2\pi)^3} \, W(\K) \, W'(\K)^*
 \label{eq:k_approx}
\end{equation}
Note that the exact relation (\ref{eq:k_mixing}) between $\hP_{\phi\phi'}(k)$ and $P_{\phi\phi'}(k')$ involves a convolution which mixes $(k,k')$, whereas the approximation in (\ref{eq:k_approx}) treats the relation as an overall constant.
This approximation suffices for purposes of this paper, where normalization is just used for plotting.

Now we specialize this general construction to the case of galaxy and velocity fields.
Consider the fields $\rho_g(\x)$ and $\hv_r(\x)$ defined in Eq.\ (\ref{eq:photoz_fields}).
These fields are related to the fields $\delta_g(\x)$ and $v_r^{\rm true}(\x)$ by:
\begin{equation}
\rho_g(\x) = W_g(\x) \delta_g(\x)
  \hspace{1.5cm}
\hv_r(\x) = W_v(\x) v_r^{\rm true}(\x)
\end{equation}
where the weight functions $W_g$, $W_v$ defined by this equation can be approximated (using Eq.\ (\ref{eq:vrec_bias})) as sums over the random catalog: 
\begin{equation} 
W_g(\x) = \frac{N_g}{N_r} \sum_{j\in\rm rand} W_j^g \delta^3(\x-\x_j)
  \hspace{1.5cm}
W_v(\x) = \frac{N_g}{N_r} \sum_{j\in\rm rand} W_j^v B(\x_j) \delta^3(\x-\x_j)
 \label{eq:Wgv}
\end{equation}
Our power spectrum normalizations $\N_{gg}$, $\N_{gv}$ are obtained by specializing Eq.\ (\ref{eq:k_approx}) above to the window functions $W_g$, $W_v$.
One more implementation detail: rather than integrating over all $\K$ as in Eq.\ (\ref{eq:k_approx}), we integrate over two $K$-shells and subtract:
\begin{align}
\N_{gg} &= \left( \int_{K < K_0} - \int_{K_0 < k < 2^{1/3}K_0} \right) 
\frac{d^3\K}{(2\pi)^3} \,
W_g(\K) W_g(\K)^* \nn \\
\N_{gv} &= \left( \int_{K < K_0} - \int_{K_0 < k < 2^{1/3}K_0} \right) 
\frac{d^3\K}{(2\pi)^3} \,
W_g(\K) W_v(\K)^* 
\end{align}
where $K_0 = 0.1$ Mpc$^{-1}$.
We include this subtraction step in order to cancel shot noise, which arises because the window functions have been approximated by sums (\ref{eq:Wgv}) over the random catalog.

\section{More on surrogate fields}
\label{app:surrogates}

In the main text, we claimed without proof that the galaxy density field $\rho_g(\x)$ has the same field-level covariance as the surrogate field $S_g(\x)$ that we constructed in \S\ref{sec:high_level_pipeline}.
We made this claim twice, once in \S\ref{ssec:surr1} assuming spectroscopic redshifts, and again in \S\ref{ssec:photoz} with a more general construction that allows photometric redshifts.
In this appendix, we will prove this statement rigorously, in the photometric case (which implies the spectroscopic case by setting $z_{\rm obs}=z_{\rm true}$ and $\sigma_z=0$).

\subsection{Precise statement of theorem}

It will be convenient to use an abstract notation, where $\y$ denotes the vector of parameters associated with each galaxy:
\begin{equation}
\y \equiv (\th, z_{\rm true}, z_{\rm obs}, \sigma_z)
\end{equation}
Let $\x_{\rm true}$, $\x_{\rm obs}$ denote the 3-d locations at sky location $\th$ and redshifts $z_{\rm true}$, $z_{\rm obs}$ respectively:
\begin{equation}
\x_{\rm true} \equiv \chi(z_{\rm true}) \, \th
  \hspace{1.5cm}
\x_{\rm obs} \equiv \chi(z_{\rm obs}) \, \th
\end{equation}
where $\chi(z)$ denotes the comoving distance to redshift $z$.
Let $P_0(\y) = P_0(\th, z_{\rm true}, z_{\rm obs}, \sigma_z)$ be the probability distribution of the randoms, including the angular mask, normalized so that:\footnote{Note that in Eq.\ (\ref{eq:p0_normalized}) and throughout this appendix, it will be convenient to use ``five-dimensional'' notation, e.g.\ $P_0$ is represented as a probability distribution on a five-dimensional space.
In practice, $P_0$ will factorize as:
\begin{equation}
P_0(\th, z_{\rm true}, z_{\rm obs}, \sigma_z)
 = \underbrace{M(\th)}_{\rm angular\ mask} \times 
 \underbrace{P(z_{\rm true}, z_{\rm obs}, \sigma_z)}_{\rm redshift\ distribution}
\end{equation}
but the results in Appendix \ref{app:surrogates} don't assume this factorization.}
\begin{equation}
\int d^5\y \, P_0(\y) = 1
  \hspace{1.5cm} \mbox{where } 
  d^5\y \equiv d^2\th \, dz_{\rm true} \, dz_{\rm obs} \, d\sigma_z
  \label{eq:p0_normalized}
\end{equation}
Let $W(\y)$ be an arbitrary ``weight function''.
We are interested in the case where $W(\y)$ is the quantity $W_i^g$ defined in Eq.\ (\ref{eq:wfkp_photoz}), but the results of this appendix will apply to an arbitrary function of $\y$.

Let $\delta_G(\x_{\rm true})$ be the following Gaussian field, defined previously in Eq.\ (\ref{eq:deltag_model2}):
\begin{equation}
\delta_G(\k,z) =
\left( b_g(z) + f_{\rm NL} \frac{2\delta_c (b_g(z)-1)}{\alpha(k,z)} \right)
\delta_{\rm lin}(\k,z)\,.
\end{equation}
We will denote $\delta_G$ either in three-dimensional notation $\delta_G(\x_{\rm true})$, or five-dimensional notation $\delta_G(\y)$.
(Note that $\delta_G$ can be viewed as a function of $\y$ which depends on $(\th,z_{\rm true})$ but not $(z_{\rm obs}, \sigma_z)$.)

Given the above definitions of $(\y, P_0, W, \delta_G)$, we now define two random fields, a ``galaxy'' field $\rho_g(\x)$, and a ``surrogate'' field $S_g(\x)$.

{\bf Definition of galaxy field $\rho_g$.}
The random field $\rho_g(\x)$ is defined by the following simulation procedure.
First, we simulate a random realization of the Gaussian field $\delta_G(\y)$.
Second, in each small volume $d^5\y$, we place a galaxy with probability:
\begin{equation}
\bar N_g P_0(\y) \big[ 1 + \delta_G(\y) \big] \, d^5\y
\end{equation}
Note that the survey geometry is incorporated in this step, since $P_0(\y)$ includes the angular mask.
Similarly, we create an unclustered ``random'' catalog by placing a random in each small volume $d^5\y$ with probability:
\begin{equation}
\bar N_r P_0(\y) \, d^5\y
\label{eq:gal_placement_probability}
\end{equation}
where $\bar N_r \gg \bar N_g$.
Third, define $\rho_g(\x)$ by summing over all galaxies with weight $W(\y)$:
\begin{equation}
\rho_g(\x) \equiv 
\bigg( \sum_{i\in \rm gal} W(\y_i) \delta^3(\x-\x_i^{\rm obs}) \bigg)
 - \frac{\bN_g}{\bN_r} 
\bigg( \sum_{j\in \rm rand} W(\y_j) \delta^3(\x-\x_j^{\rm obs}) \bigg)
\label{eq:app_rhog_def}
\end{equation}
{\bf Definition of surrogate field $S_g$.}
Similarly, the random field $S_g(\x)$ is defined by the following simulation procedure.
First, we simulate a random realization of the Gaussian field $\delta_G(\y)$.
Second, in each small volume $d^5\y$, place a galaxy with probability:
\begin{equation}
\bar N_r P_0(\y) \, d^5\y
\label{eq:rand_placement_probability}
\end{equation}
Third, we define $S_g(\y)$ by ``painting'' the random field $\delta_G$ onto the random catalog, and adding Poisson noise by hand:
\begin{equation}
S_g(\x) \equiv \frac{\bar N_g}{\bar N_r} \sum_{j\in\rm rand} W(\y_j) \big( \delta_G(\x_j^{\rm true}) + \eta_j \big) \delta^3(\x-\x_j^{\rm obs})
\label{eq:app_Sg_def}
\end{equation}
where $\eta_j$ is an uncorrelated Gaussian random variable with:
\begin{equation}
\big\langle \eta_j \eta_k \big\rangle
  = \left( \frac{\bar N_r}{\bar N_g} - \big\langle \delta_G(\x_j^{\rm true})^2 \big\rangle \right) \delta_{jk}
  \label{eq:app_eta_eta}
\end{equation}
{\bf Statement of theorem.}
In this appendix, we will prove the following theorem: {\bf the galaxy field $\rho_g(\x)$ and the surrogate field $S_g(\x)$ defined in Eqs.\ (\ref{eq:app_rhog_def}), (\ref{eq:app_Sg_def}) have the same field-level covariance:}
\begin{equation}
\big\langle \rho_g(\x) \rho_g(\x') \big\rangle = 
\big\langle S_g(\x) S_g(\x') \big\rangle
  \label{eq:main_theorem}
\end{equation}
We emphasize that in Eq.\ (\ref{eq:main_theorem}) we are considering the full field-level covariance $\langle \rho(\x) \rho(\x') \rangle$ as a function of $(\x,\x')$ with no angle-averaging.
Thus, the theorem shows that the surrogate fields fully capture effects such as survey geometry, redshift dependence, and photo-$z$ errors.

\subsection{Strategy of proof}
\label{ssec:strategy_of_proof}

We define fields $\trho(\y), \tS_g(\y)$ on the five-dimensional space $\y = (\th, z_{\rm true}, z_{\rm obs}, \sigma_z)$ by slightly modifying the definitions (\ref{eq:app_rhog_def}), (\ref{eq:app_Sg_def}) of $\rho_g(\x), S_g(\x)$ as follows:
\begin{align}
\trho_g(\y) &\equiv 
\bigg( \sum_{i\in \rm gal} W(\y_i) \delta^5(\y-\y_i) \bigg)
 - \frac{\bN_g}{\bN_r} 
\bigg( \sum_{j\in \rm rand} W(\y_j) \delta^5(\y-\y_j) \bigg)
 \nn \\
\tS_g(\y) &\equiv \frac{\bar N_g}{\bar N_r} \sum_{j\in\rm rand} W(\y_j) \big( \delta_G(\y_j) + \eta_j \big) \delta^5(\y-\y_j)
\end{align}
In \S\ref{ssec:rhog_2pcf}, \S\ref{ssec:Sg_2pcf}, we will show that the 5-d fields $\trho_g(\y)$, $\tS_g(\y)$ have the same covariance:
\begin{equation}
\big\langle \trho_g(\y) \trho_g(\y') \big\rangle = 
\big\langle \tS_g(\y) \tS_g(\y') \big\rangle
\label{eq:main_theorem_5d}
\end{equation}
This implies our main theorem (\ref{eq:main_theorem}), that the 3-d fields $\rho_g(\x), S_g(\x)$ have the same covariance.
This statement is not obvious, but follows from the following formal argument.

The 3-d fields $\rho_g(\x)$ and $S_g(\x)$ are related to the 5-d fields $\trho_g(\x)$ and $\tS_g(\x)$ via a formal process of ``integrating out'' the variables $(z_{\rm true}, \sigma_z)$:
\begin{align}
\rho_g(\x) = \int d^5\y\, \trho_g(\y) 
\delta^3\big( \x - \chi(z_{\rm obs}\th) \big)
 \hspace{1.5cm}
S_g(\x) = \int d^5\y\, \tS_g(\y) 
\delta^3\big( \x - \chi(z_{\rm obs}\th) \big)
\label{eq:5d_to_3d}
\end{align}
where $\y = (\th,z_{\rm true},z_{\rm obs},\sigma_z)$ on the RHS.
Generally speaking, if two random variables $(\trho_g, \tS_g)$ have the same covariance, and the same linear operation (\ref{eq:5d_to_3d}) is applied to both, then the resulting random variables $(\rho_g, S_g)$ also have the same covariance.
Therefore Eq.\ (\ref{eq:main_theorem_5d}) implies our main theorem (\ref{eq:main_theorem}).

\subsection{Computing $\langle \tilde\rho_g({\bf y}) \tilde\rho_g({\bf y}') \rangle$}
\label{ssec:rhog_2pcf}

In this section we compute the LHS $\langle \trho_g(\y) \trho_g(\y') \rangle$ of Eq.\ (\ref{eq:main_theorem_5d}).
We will compute the expectation value in two steps.
First, we will fix a random realization of $\delta_G(\y)$ and compute the expectation value $\langle \cdot \rangle_{\rm pl}$ over random placements of the galaxies $\{ \y_i \}$.
Second, we will take an ``outer'' expectation value over random realizations of $\delta_G$.

We will assume $\bN_r \gg \bN_g$.
In this limit, the field $\trho_g(\y)$ can be written:
\begin{equation}
\trho_g(\y) = \bigg( \sum_{i\in \rm gal} W(\y_i) \delta^5(\y-\y_i) \bigg)
 - \bN_g W(\y)  P_0(\y)  
\end{equation}
Now we take the expectation value $\langle \cdot \rangle_{\rm pl}$ over galaxy placements, using the placement probability (\ref{eq:gal_placement_probability}).
The expectation value  is the sum of ``2-halo'' and ``1-halo'' terms:
\begin{align}
\big\langle \trho_g(\y) \trho_g(\y') \big\rangle_{\rm pl}
  &= \zeta_{2h}(\y,\y') + \zeta_{1h}(\y,\y') 
  \label{eq:rho_1h_2h} \\ 
\zeta_{2h}(\y,\y') 
   &= \bar N_g^2 W(\y) W(\y') P_0(\y) P_0(\y')
   \Big[ 
   \big( 1 + \delta_G(\y) \big)
   \big( 1 + \delta_G(\y') \big) 
   - 1 \Big] \nn \\
\zeta_{1h}(\y,\y') &=
  \bar N_g W(\y)^2 P_0(\y) \big( 1 + \delta_G(\y) \big) \delta^5(\y-\y') \nn
\end{align}
We omit details of this calculation, since it is so similar to the standard halo model.
Next, we take the expectation value of Eq.\ (\ref{eq:rho_1h_2h}) over random realizations of $\delta_G$:
\begin{equation}
\big\langle \trho_g(\y) \trho_g(\y') \big\rangle =
W(\y) W(\y') \Big[ 
 \bN_g^2 P_0(\y) P_0(\y') C(\y,\y') 
  + \bN_g P_0(\y) \delta^5(\y-\y')
\label{eq:app_rho_rho}
\Big]
\end{equation}
where $C(\y,\y')$ denotes the covariance of the Gaussian random field $\delta_G(\y)$:
\begin{equation}
C(\y,\y') \equiv \big\langle \delta_G(\y) \delta_G(\y') \big\rangle
\label{eq:C_def}
\end{equation}

\subsection{Computing $\langle \tilde S_g({\bf y}) \tilde S_g({\bf y}') \rangle$}
\label{ssec:Sg_2pcf}

In this section we compute the RHS $\langle \tS_g(\y) \tS_g(\y') \rangle$ of Eq.\ (\ref{eq:main_theorem_5d}).
We write $\tS_g(\y)$ in the form:
\begin{equation}
\tS_g(\y) = \sum_j w_j \delta^5(\y-\y_j)
  \hspace{1.5cm} \mbox{where }
w_j \equiv \frac{\bar N_g}{\bar N_r} \,
  W(\y_j) \, 
  \big( \delta_G(\y_j) + \eta_j \big)
\label{eq:wj_def}
\end{equation}
We will compute the expectation value $\langle \tS_g(\y) \tS_g(\y') \rangle$ in two steps.
First, we will fix a realization of the random catalog $\{ \y_j \}$, and compute the expectation value $\langle \cdot \rangle_w$ over the random variable $w_j$.
Second, we will take an ``outer'' expectation value over the random catalog $\{ \y_j \}$.
The first step can be done as follows:
\begin{align}
\big\langle \tS_g(\y) \tS_g(\y') \big\rangle_w
  &= \sum_{jk} \big\langle w_j w_k \big\rangle \,
      \delta^5(\y-\y_j) \delta^5(\y'-\y_k) \nn \\
  &= \sum_{jk} \frac{\bN_g^2}{\bN_r^2} \,
    W(\y_j) W(\y_k)
  \left[ C(\y_j,\y_k) + 
  \left( \frac{\bN_r}{\bN_g} - C(\y_j,\y_j) \right) \delta_{jk}
  \right]
  \delta^5(\y-\y_j) \delta^5(\y'-\y_k) \nn \\
 &= \frac{\bN_g^2}{\bN_r^2} W(\y) W(\y') 
 \bigg[ C(\y,\y') \sum_{jk} \delta^5(\y-\y_j) \delta^5(\y'-\y_k) \nn \\
 & \hspace{3cm}
   + \bigg( \frac{\bN_r}{\bN_g} - C(\y,\y) \bigg)
     \delta^5(\y-\y') \sum_j \delta^5(\y-\y_j) \bigg]
\label{eq:Sg_precursor}
\end{align}
where in the second line, we have computed the expectation value $\langle w_j w_k \rangle$ using the definition (\ref{eq:wj_def}) of $w_j$, and Eq.\ (\ref{eq:app_eta_eta}) for $\langle \eta_j \eta_k \rangle$.
The covariance $C(\y,\y')$ was defined in Eq.\ (\ref{eq:C_def}).

The next step is to take the ``outer'' expectation value over the random catalog $\{ \y_j \}$.
Using Eq.\ (\ref{eq:rand_placement_probability}) for the placement probability, one can compute the following expectation values:
\begin{align}
\bigg\langle \sum_j \delta^5(\y-\y_k) \bigg\rangle
 &= \bN_r P_0(\y) \nn \\
\bigg\langle \sum_{jk} \delta^5(\y-\y_j) \delta^5(\y'-\y_k) \bigg\rangle
 &= \bN_r^2 P_0(\y) P_0(\y') + \bar N_r P_0(\y) \delta^5(\y-\y')
\label{eq:Sg_rcat_ev}
\end{align}
where the RHS of the second line is the sum of ``2-halo'' and ``1-halo'' terms.
Plugging Eq.\ (\ref{eq:Sg_rcat_ev}) into Eq.\ (\ref{eq:Sg_precursor}) and simplifying, we get:
\begin{equation}
\big\langle \tS_g(\y) \tS_g(\y') \big\rangle
 = W(\y) W(\y') \Big[ 
   \bN_g^2 P_0(\y) P_0(\y') C(\y,\y')
     + \bN_g P_0(\y) \delta^5(\y-\y')
 \Big]
\end{equation}
Comparing to our earlier result (\ref{eq:app_rho_rho}) for $\langle \trho_g(\y) \trho_g(\y') \rangle$, we see that the two are equal, i.e.
\begin{equation}
\big\langle \trho_g(\y) \trho_g(\y') \big\rangle = 
\big\langle \tS_g(\y) \tS_g(\y') \big\rangle
\end{equation}
As explained in \S\ref{ssec:strategy_of_proof}, this implies the main theorem of this appendix (Eq.\ (\ref{eq:main_theorem})).

\section{Validating Surrogate Fields with SDSS Mocks}
\label{sec:Appendix_SDSSvalidation}

\begin{figure}
    \centering
    \includegraphics[width=0.65\textwidth]{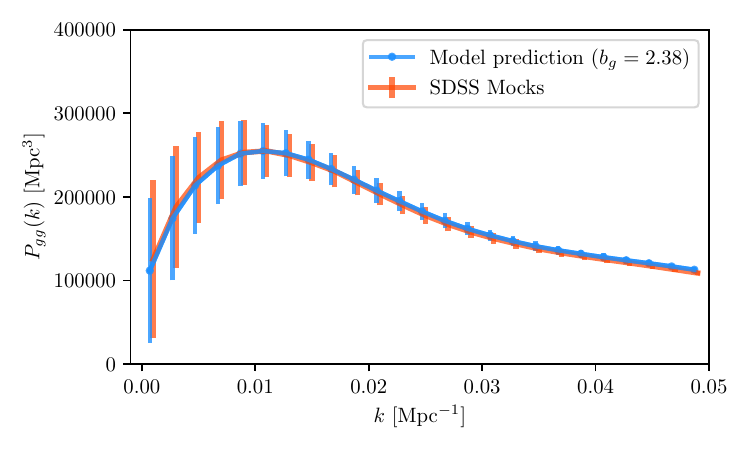}
    \vspace*{-0.5cm}
    \caption{Comparison between mean power spectra and error bars from SDSS surrogate simulations (blue) and SDSS mocks (orange). We find a good match between both mean power spectra and their error bars.}
    \label{fig:validation_with_sdss_powspec}
\end{figure}

\begin{figure}
    \centering
    \includegraphics[width=0.8\textwidth]{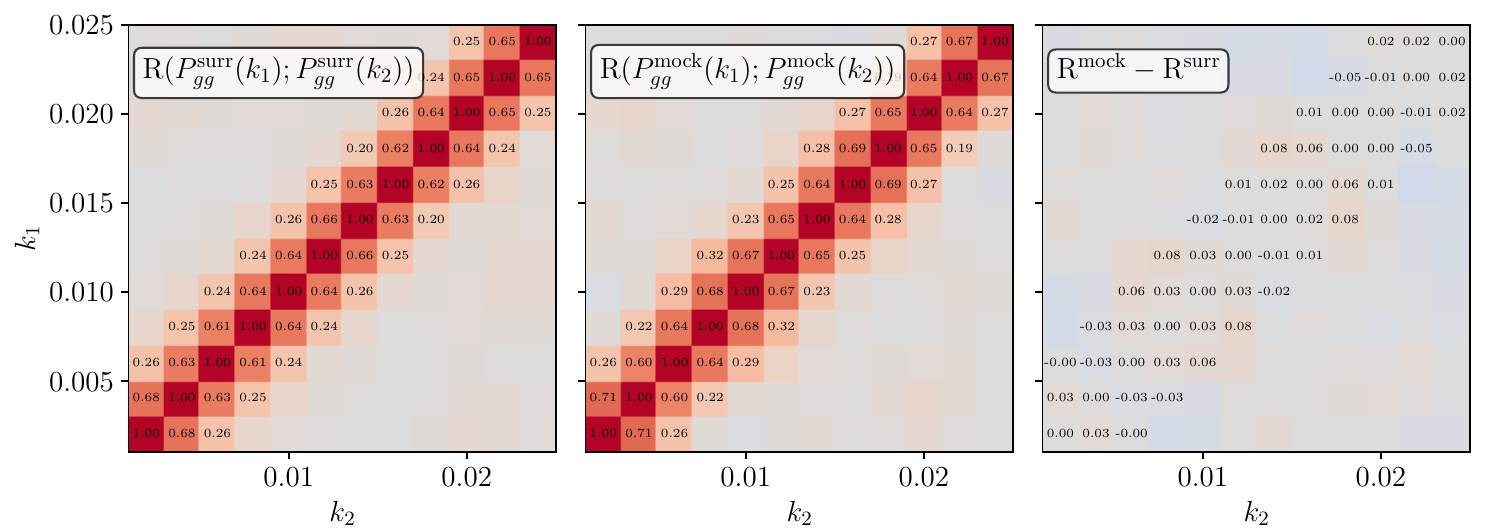}\caption{Comparison between the correlation matrices calculated from SDSS mocks and SDSS surrogate simulations. We find an excellent match between the two. The correlation matrix is defined in Eq.~\eqref{eq:rmatrix}.}
    \label{fig:validation_with_sdss}
\end{figure}

As explained in \S\ref{ssec:surr1}, we use surrogate simulations for two purposes: (1) to predict estimated power spectra (e.g.\ $\smash{\langle \hP_{gv}(k) \rangle}$) in a given model, and (2) to assign error bars to power spectra.
To show that (1) is valid, we have given analytic arguments (Appendices \ref{app:surrogates}, \ref{app:hvr_norm}) that the surrogate simulations correctly model the field-level covariance $\langle S(\x) S(\x') \rangle$, including anisotropic effects such as the survey window function.
To show that (2) is valid, we argued in \S\ref{ssec:surr1} that we are mainly interested in large scales, where fields are close to Gaussian, and we expect the field-level covariance to predict the power spectrum covariance (via Wick's theorem).
However, it would also be valuable to have a direct comparison of the error bars inferred from surrogate simulations and mock catalogs.

In this appendix, we compare the power spectrum covariance $\mbox{Cov}(P_{gg}(k), P_{gg}(k'))$ from surrogate simulations and mocks.
Since mocks are not available for DESILS, we used the Sloan Digital Sky Survey Baryon Oscillation Spectroscopic Survey (SDSS BOSS DR12~\citep{BOSS:2016wmc, 2013AJ....145...10D}) data products for this comparison.

We use 1000 SDSS mock catalogs to compute the galaxy auto-power spectrum and its covariance. The mean and variance from these mocks are shown in orange in Fig.~\ref{fig:validation_with_sdss_powspec}. The mock spectra are generated using a mocks-minus-randoms approach, with randoms matched to the SDSS simulations. For comparison, we also show the mean and variance of power spectra from 1500 surrogate simulations (in blue), constructed using SDSS-matched randoms and assuming a galaxy bias of $b_g = 2.38$. We find good agreement in both the mean and variance between the mocks and surrogate spectra.

In Fig.~\ref{fig:validation_with_sdss} we compare the correlation matrix of SDSS mocks and our surrogate simulations. We defined the correlation matrix in Eq.~\eqref{eq:rmatrix}. The left and middle panels correspond to rotation matrices corresponding to surrogate field simulations and mocks respectively. The right panel shows the difference between the two matrices. We find comparing the correlation matrices that our surrogate field method provides an excellent match to the mocks simulations with differences between adjacent power spectra $k$-bin covariance remaining well below 10 percent.  

\section{Normalization of the velocity reconstruction field $\hv_r(\x)$}
\label{app:hvr_norm}

\subsection{Statement of main result}

Recall the definition (\ref{eq:photoz_fields}) of the velocity reconstruction estimator:
\begin{equation}
\hv_r(\x) = \sum_{i\in\rm gal} W_i^v \, \tT(\th_i) \, \delta^3(\x-\x_i^{\rm obs})
\end{equation}
where we have omitted mean-subtraction for simplicity, and assumed photometric redshifts for generality.
We are interested in the normalization of the velocity reconstruction, i.e.\ the relation between $\hv_r(\x)$ and the true radial velocity field $v_r(\x)$.

In this appendix, we define a field $\tS(\x)$ by:
\begin{equation}
\tS(\x) = \frac{N_g}{N_r} \sum_{j\in\rm rand} W_j^v \, W_{\rm CMB}(\th_j) \, \tB_0(z_j^{\rm true}) \, v_r(\x_j^{\rm true})
\end{equation}
where $v_r(\x)$ is the true (not simulated) large-scale radial velocity field, and $\tB_0(z)$ is defined by:
\begin{equation}
\tB_0(z) \equiv \frac{K(z)}{\chi(z)^2}
\int \frac{d^2\L}{(2\pi)^2} \, b_L \, F_L \, P_{ge}^{\rm true}(k,z)_{k=L/\chi(z)}
\label{eq:tB_def}
\end{equation}
Note that $\tS(\x)$ is very similar to our surrogate field $S_v^{\rm sig}(\x)$ defined in Eq.\ (\ref{eq:Sv_sig_photo}), with two differences.
First, $\tS$ is defined using the true $v_r$, whereas $S_v^{\rm sig}$ is defined with a simulated $v_r$.
Second, $\tS$ uses the true galaxy-electron power spectrum $P_{ge}^{\rm true}(k_S)$ (via Eq.\ (\ref{eq:tB_def})), whereas $S_v$ uses a fiducial $P_{ge}^{\rm fid}(k_S)$ (via Eq.\ (\ref{eq:B_def})).

The goal of this appendix is to prove the following ``main result'':
\begin{equation}
\big\langle \hv_r(\x) \big\rangle \approx \tS(\x)
\label{eq:app_main_result}
\end{equation}
where in this equation (and throughout this appendix):
\begin{itemize}
\item 
The notation $\langle \cdot \rangle$ denotes an average over small-scale modes $\delta_g(\k_S), \delta_e(\k_S)$, in a {\bf fixed realization of the long-wavelength velocity field $v_r(\k_L)$.}
\item
The notation $\approx$ means that the LHS and RHS are equal in the limit $N_r \gg N_g$ of a large random catalog, and on large scales. (That is, after taking a Fourier transform $\int d^3\x \, e^{i\k_L\cdot\x} (\cdots)$ where $k_L$ is a small wavenumber compared to a kSZ scale $k_S \sim 1$ Mpc$^{-1}$.)
\end{itemize}
Our main result (\ref{eq:app_main_result}) implies several statements in the body of the paper:

\begin{itemize}

\item 
In \S\ref{ssec:hvr}, we claimed (Eq.\ \ref{eq:vrec_bias}) that in a spectroscopic survey, the normalization of the velocity reconstruction is:
\begin{equation}
\big\langle \hat v_r(\x) \big\rangle = b_v(z) \, \bar n_v(\x) \, W_{\rm CMB}(\th) \, B_0(z) \, v_r^{\rm true}(\x)
\label{eq:app_vr_norm}
\end{equation}
where $\bar n_v$, $B_0$, $b_v$ were defined in Eqs.\ (\ref{eq:nv_def}), (\ref{eq:B_def}), (\ref{eq:bv_def}).

To see that this follows from our main result (\ref{eq:app_main_result}), we note that for any function $f(\x)$, we have:
\begin{equation}
\lim_{N_r\rightarrow\infty} \frac{N_g}{N_r} \sum_{j\in\rm rand} W_j^v f(\x) = n_v(\x) f(\x)
\end{equation}
This follows from the definition (\ref{eq:nv_def}) of $n_v(\x)$.
Similarly, by unraveling definitions (Eqs.\ (\ref{eq:B_def}), (\ref{eq:bv_def}), (\ref{eq:tB_def})) we see that $\tB_0(z) = b_v(z) B_0(z)$.
Making these substitutions in our main result (\ref{eq:app_main_result}), we obtain Eq.\ (\ref{eq:app_vr_norm}) above.

Note that our main result (\ref{eq:app_main_result}) can be viewed as a generalizing the $\hat v_r(\x)$ normalization to a photometric survey (where we have chosen to write the RHS as a sum over a random catalog).

\item
We claimed that the surrogate field $S_v^{\rm sig}(\x)$ correctly models the field-level covariance of $\hv_r(\x)$, in the limit of zero reconstruction noise (where reconstruction noise is added in a separate term $S_v^{\rm noise}(\x)$).
We made this claim twice, in \S\ref{ssec:surr2} for a spectroscopic survey, and in \S\ref{ssec:photoz} for the more general case of a photometric survey.
This follows from our main result (\ref{eq:app_main_result}), plus the observation that the surrogate field $S_v^{\rm sig}(\x)$ is related to our $\tS_0(\x)$ by randomizing the velocity field $v_r(\x)$.

\end{itemize} 

\subsection{Proof of main result}

First, we claim that it suffices to prove the main result (\ref{eq:app_main_result}) in the following special case:
\begin{enumerate}
\item The galaxy survey does not include an angular mask. 
(However, we allow $W_{\rm CMB}(\th)$ to be arbitrary, and to have structure on small scales.)
\item The per-galaxy weighting is given by $W_i^v=1$.
\item The galaxy catalog is spectroscopic, i.e.\ $\sigma_z=0$.
\end{enumerate}
To see this, note that if (\ref{eq:app_main_result}) is true without an angular mask, then it is still true after applying an angular mask on both sides (item \#1).
Similarly, we can apply a per-galaxy weighting on both sides (item \#2), or a photo-$z$ convolution on both sides (item \#3).
We will make these three assumptions in the rest of this section.

We will prove (\ref{eq:app_main_result}) in a ``snapshot'' geometry, where the universe is a 3-d periodic box at a fixed time, and the CMB is a 2-d flat-sky field along one face of the box.
The snapshot geometry is an approximation of the true ``lightcone'' geometry, and neglects sky curvature and redshift evolution.
However, Eq.\ (\ref{eq:app_main_result}) does not involve correlation functions on large scales (only small-scale quantities such as $P_{ge}(k_S)$ or the small-scale CMB filter $F_L$), so if (\ref{eq:app_main_result}) is true for a snapshot geometry, it must be true in a lightcone geometry as well.

Our notation for the snapshot geometry is as follows.
Slowly evolving quantities, such as $\chi(z)$ or $K(z)$, are evaluated at a fixed time, and denoted $\chi_*$ or $K_*$.
We denote 3-d positions by $\x = (\x_\perp, x_\parallel)$.
We identify 2-d positions $\th$ with transverse 3-d positions $\x_\perp = \chi_* \th$.

As an example of the snapshot geometry notation, the kSZ line-of-sight integral (\ref{eq:tksz_line_of_sight}) can be written:
\begin{equation}
T_{\rm kSZ}(\th)
 = K_* \int dx_\parallel \, v_r(\x) \, \delta_e(\x) \Big|_{\x=(\chi_*\th, x_\parallel)}
 \label{eq:tksz_snapshot}
\end{equation}
Since the 3-d galaxy number density $n_g(z)$ is slowly varying in $z$, we will treat it as a constant $n_*$ in the snapshot geometry.
The velocity reconstruction $\hv_r(\x)$ can be written:
\begin{equation}
\hv_r(\x) = n_* \, \delta_g(\x) \, \tT(\th) \big|_{\th=\x_\perp/\chi_*}
\label{eq:hvr_snapshot}
\end{equation}
The main result (\ref{eq:app_main_result}) that we want to prove can be written:
\begin{equation}
\big\langle \hv_r(\x) \big\rangle = n_* B_* \, 
v_r(\x) \, W_{\rm CMB}(\th)\big|_{\th=\x_\perp/\chi_*}
\hspace{1cm} \mbox{where }
B_* \equiv \frac{K_*}{\chi_*^2} \int \frac{d^2\L}{(2\pi)^2} \, b_L F_L P_{ge}(k)_{k=L/\chi_*} 
\label{eq:app_hvr_goal}
\end{equation}
In the rest of this appendix, we will derive Eq.\ (\ref{eq:app_hvr_goal}), by direct calculation in the snapshot geometry.
It will be convenient to define real-space counterparts of the quantities $P_{ge}(k)$, $b_L$, $F_L$:
\begin{equation}
\zeta_{ge}(\r) \equiv \int \frac{d^3\k}{(2\pi)^3} \, P_{ge}(k) e^{i\k\cdot\r}
 \hspace{1cm}
b(\Th) \equiv \int \frac{d^2\L}{(2\pi)^2} \, b_L e^{i\L\cdot\Th}
 \hspace{1cm}
f(\Th) \equiv \int \frac{d^2\L}{(2\pi)^2} \, F_L e^{i\L\cdot\Th}
\label{eq:zeta_ge_def}
\end{equation}
Using this notation, the filtered CMB $\tT(\th)$ defined in Eq.~(\ref{eq:tT_allsky}) can be written in terms of the observed CMB $T_{\rm obs}(\th)$ as:
\begin{equation}
\tT(\th) = \int d^2\th \, f(\th-\th') W_{\rm CMB}(\th') T_{\rm obs}(\th')
\end{equation}
Similarly, the observed CMB $T_{\rm obs}(\th')$ contains the kSZ temperature $T_{\rm kSZ}(\th)$, with a beam convolution:
\begin{equation}
T_{\rm obs}(\th') \supset \int d^2\th'' \, b(\th'-\th'') T_{\rm kSZ}(\th'')
\end{equation}
where the notation ``$\supset$'' means we have omitted non-kSZ terms (e.g.\ primary CMB, instrumental noise) on the RHS.
Combining the previous two equations with Eqs.\  (\ref{eq:tksz_snapshot}), (\ref{eq:hvr_snapshot}), we can write the following relation between $\hv_r(\x)$ and the large-scale structure fields $v_r, \delta_e$:
\begin{equation}
\hv_r(\x) = n_* K_* \delta_g(\x) 
\int d^2\th' \, d^2\th'' \, dx''_\parallel \, 
f(\th-\th') \, W_{\rm CMB}(\th') \, b(\th'-\th'') \, v_r(\x'') \, \delta_e(\x'')
\Big|_{\substack{\th=\x_\perp/\chi_* \\ \x''=(\chi_*\th'', x''_\parallel)}}
\end{equation}
Now we can take the expectation value (in a fixed realization of $v_r(\x)$), by replacing $\delta_g(\x) \delta_e(\x'') \rightarrow \zeta_{ge}(\x-\x'')$ on the RHS (where $\zeta_{ge}$ was defined in Eq.~(\ref{eq:zeta_ge_def})):
\begin{equation}
\big\langle \hv_r(\x) \big\rangle
 \supset n_* K_* 
\int d^2\th' \, d^2\th'' \, dx''_\parallel \, 
f(\th-\th') \, W_{\rm CMB}(\th') \, b(\th'-\th'') \, v_r(\x'') \, \zeta_{ge}(\x-\x'')
\Big|_{\substack{\th=\x_\perp/\chi_* \\ \x''=(\chi_*\th'', x''_\parallel)}}
\end{equation}
We make the approximation that the radial velocity field $v_r(\x'')$ is slowly varying on kSZ scales ($k_S^{-1} \sim 1$ Mpc), so that we can approximate $v_r(\x'') \approx v_r(\x)$ and pull it out of the integral:
\begin{equation}
\big\langle \hv_r(\x) \big\rangle
 = n_* K_* \, v_r(\x) 
\int d^2\th' \, d^2\th'' \, dx''_\parallel \, 
f(\th-\th') \, W_{\rm CMB}(\th') \, b(\th'-\th'')  \, \zeta_{ge}(\x-\x'')
\Big|_{\substack{\th=\x_\perp/\chi_* \\ \x''=(\chi_*\th'', x''_\parallel)}}
\label{eq:app_hvr_precursor}
\end{equation}
Now consider the following subexpression:
\begin{equation}
I \equiv \int d^2\th'' \, dx''_\parallel \, b(\th'-\th'') \, \zeta_{ge}(\x-\x'')
\Big|_{\substack{\x''=(\chi_*\th'', x''_\parallel)}}
\end{equation}
After a short calculation using the defintions of $b(\Theta)$ and $\zeta_{ge}(r)$ in Eq.\ (\ref{eq:zeta_ge_def}), the integral can be done exactly:
\begin{equation}
I = \frac{1}{\chi_*^2} \, \eta(\th-\th')\big|_{\th=\x_\perp/\chi_*}
  \hspace{1cm} \mbox{where }
\eta(\Th) \equiv \int \frac{d^2\L}{(2\pi)^2} \, 
  e^{-i\L\cdot\Th} \,
  b_L P_{ge}(k)_{k=L/\chi_*}
  \label{eq:eta_def}
\end{equation}
Plugging this expression for $I$ back into (\ref{eq:app_hvr_precursor}), we get:
\begin{equation}
\big\langle \hv_r(\x) \big\rangle
 = \frac{n_* K_*}{\chi_*^2} \, v_r(\x) \, \int d^2\th' \, \rho(\th-\th') W_{\rm CMB}(\th')
 \Big|_{\th=\x_\perp/\chi_*}
\hspace{1cm} \mbox{where }
  \rho(\Th) \equiv f(\Th) \eta(\Th)
  \label{eq:rho_def}
\end{equation}
The convolution kernel $\rho(\th-\th')$ is narrowly peaked, i.e.\ its Fourier transform $\rho(\L)$ is approximately constant (in $L$) on large scales compared to the kSZ ($L \ll 4000$).
Since we are only interested in such large scales, we can make the approximation:
\begin{equation}
\big\langle \hv_r(\x) \big\rangle
 \approx \frac{n_* K_*}{\chi_*^2} \rho_0 \, v_r(\x) \, W_{\rm CMB}(\th)
 \Big|_{\th=\x_\perp/\chi_*}
\hspace{1cm} \mbox{where }
\rho_0 \equiv \int d^2\Th\, \rho(\Th)
\end{equation}
A short calculation, unraveling definitions of $f(\Th)$, $\eta(\Th)$ and $\rho(\Th)$ in Eqs.\ (\ref{eq:zeta_ge_def}), (\ref{eq:eta_def}), (\ref{eq:rho_def}), shows that $\rho_0$ is given by:
\begin{equation}
\rho_0 = \int \frac{d^2\L}{(2\pi)^2} \, b_L F_L P_{ge}(k)_{k=L/\chi_*}  
\end{equation}
Plugging this in to the previous equation and using the definition (\ref{eq:app_hvr_goal}) of $B_*$, we get:
\begin{equation}
\big\langle \hv_r(\x) \big\rangle = n_* B_* \, 
v_r(\x) \, W_{\rm CMB}(\th)\big|_{\th=\x_\perp/\chi_*}
\end{equation}
as desired (Eq.\ (\ref{eq:app_hvr_goal})).

\end{document}